\documentclass[twocolumn,a4paper,10pt,showpacs,aps,pra,floatfix]{revtex4-1}

\usepackage[utf8]{inputenc} 
\usepackage[english]{babel} 
\usepackage{csquotes} 
\usepackage{microtype} 
\usepackage{textcomp} 
\usepackage{siunitx} 

\newcommand{\linkcolor}{blue} \usepackage{hyperref} 
\hypersetup{colorlinks=true, %
  linkcolor=\linkcolor, %
  citecolor=\linkcolor, %
  filecolor=\linkcolor, %
  urlcolor=\linkcolor, %
  pdftitle={Quantum interference of a single spin excitation with a
    macroscopic atomic ensemble}, %
  pdfauthor={S. L. Christensen, J.-B. Béguin, E. Bookjans,
    H. L. Sørensen, J. H. Müller, J. Appel, E. S. Polzik}, %
  pdfsubject={Quantum Optics, Quantum noise, Quantum state engineering
    and measurements}, %
  pdfkeywords={Quantum Optics, Quantum noise, Quantum state
    engineering and measurements} %
} %

\usepackage{color} %
\usepackage{graphicx} 
\graphicspath{{images/}{./}}

\usepackage{mathtools} %
\usepackage{xpatch} 

\makeatletter%
\def \@labelsection{%
  \@ifundefined{@sectioncntformat}%
  {\@seccntformat}%
  {\@sectioncntformat}{section}%
}%
\def \@labelsubsection{\@labelsection.\thesubsection}%
\def \@labelsubsubsection{\@labelsubsection.\thesubsubsection}%
\xpatchcmd{\@sect@ltx}{\@xsect}{%
  \let\@hskip\hskip%
  \def \hskip { \@hskip 0em plus}%
  \let\@MakeTextUppercase\MakeTextUppercase%
  \def \MakeTextUppercase{}%
  \edef \@currentlabelname{%
    \@hangfrom@section{}{\csname @label#1\endcsname}{#8}%
  } %
  \let\MakeTextUppercase\@MakeTextUppercase%
  \let\hskip\@hskip%
  \@xsect}{}{}
\makeatother

\DeclareMathOperator{\var}{var} %
\DeclareMathOperator{\sinc}{sinc} %
\newcommand{\Natom}{N_{\text{a}}} %
\newcommand{\La}{L_{\text{a}}} %

\newcommand{\ket}[1]{\left|#1\right\rangle} %
\newcommand{\bra}[1]{\left\langle #1\right|} %
\newcommand{\ketbra}[2]{\ket{#1}\!\!\bra{#2}} %
\newcommand{\dd}{\mathrm{d}} %
\newcommand{\e}{\mathrm{e}} %

\DeclareSIUnit{\gauss}{G}

\addto{\captionsenglish}{}

\newcommand{\fref}[2][]{Fig.~\ref{#2}\textcolor{\linkcolor}{#1}} 
\newcommand{\sref}[1]{Sec.~\nameref{#1}} 

\begin{document}
\title{Quantum interference of a single spin excitation with a
  macroscopic atomic ensemble}

\newcommand{\NBI}{QUANTOP, Niels Bohr Institute, University of
  Copenhagen, Blegdamsvej 17, 2100 Copenhagen, Denmark}
\newcommand{\correspondingauthors} { \email[Corresponding
  Authors:]{polzik@nbi.dk} \email{jappel@nbi.dk}}

\author{S. L. Christensen} \affiliation{\NBI} %
\author{J.-B. Béguin} \affiliation{\NBI} %
\author{E. Bookjans} \affiliation{\NBI} %
\author{H. L. Sørensen} \affiliation{\NBI} %
\author{J. H. Müller} \affiliation{\NBI} %
\author{J. Appel} \correspondingauthors \affiliation{\NBI} %
\author{E. S. Polzik} \correspondingauthors \affiliation{\NBI} %
\begin{abstract}
  We report on the observation of quantum interference of a collective
  single spin excitation with a spin ensemble of $\Natom \approx
  \num{e5}$ atoms.  Detection of a single photon scattered from the
  atoms creates the single spin excitation, a Fock state embedded in
  the collective spin of the ensemble.  The state of the atomic
  ensemble is then detected by a quantum nondemolition measurement of
  the collective spin.  A macroscopic difference of the order of
  $\sqrt{\Natom}$ in the marginal distribution of the collective spin
  state arises from the interference between the single excited spin
  and $\Natom$ atoms. These hybrid discrete-continuous manipulation
  and measurement procedures of collective spin states in an atomic
  ensemble pave the road towards generation of even more exotic
  ensemble states for quantum information processing, precision
  measurements, and communication.
\end{abstract}


\ifdefined\svnid { \textcolor{green}{\svnFullRevision*{\svnrev} by
    \svnFullAuthor*{\svnauthor} \newline %
    Last changed date: \svndate } } \fi

\maketitle
\section{Introduction} \label{sec:introduction}

The development of interfaces between quantum systems plays a large
role in present-day quantum information research. One of the most used
interfaces is based on the interaction between light and atomic
ensembles~\cite{Hammerer2010, Kimble2008}.  Until now, predominantly,
two different approaches based on either discrete or continuous
variables have been used. The discrete method is based on collective
single excitations, photon counting, and mapping of the atomic state
into a photonic state which is then characterized~\cite{Choi2010,
  Kimble2008, MacRae2012,Duan2001,Specht2011}. The continuous-variable
schemes use atomic homodyne measurements which allow for deterministic
protocols, such as quantum teleportation \cite{Hammerer2010,
  Krauter2013}, spin squeezing and atomic tomography~\cite{Appel2009,
  Leroux2010a,Fernholz2008}, quantum-assisted
metrology~\cite{Wasilewski2010, Louchet-Chauvet2010, Koschorreck2010},
and quantum memories~\cite{Hammerer2010,Simon2010}. A general feature
of the continuous-variable approach is its high-efficiency state
characterization and mode selectivity.  Hybrid discrete-continuous
quantum state generation has been demonstrated in pure photonic
systems~\cite{Neergaard-Nielsen2006, Ourjoumtsev2007, Takeda2013,
  Lvovsky2013, Bruno2013}. Here, we report on a hybrid
discrete-continuous protocol combining a collective atomic excitation
heralded by a single-photon count with a continuous measurement of the
atomic state \emph{directly} in the ensemble.  In combination with
quantum nondemolition (QND) measurement-induced squeezing
\cite{Appel2009}, the discrete manipulation of the excitation number
allows for creation of Schrödinger's cat states
\cite{Neergaard-Nielsen2006} within a quantum memory which are a
valued resource for quantum repeater protocols
\cite{Brask2010}. States created by this method can improve
measurements beyond the standard quantum limit \cite{McConnell2013}.
The experiment presented here unifies two main approaches to
atom-light quantum interfaces: first, a single excitation is generated
via a Raman-type process (where a direct retrieval would result in a
single photon in the output mode) \cite{Duan2001}; then, a
Faraday-type (QND) memory readout~\cite{Hammerer2010} of the resulting
interference is performed.

\section{Theory} \label{sec:theory}

Our experiment is conducted on an atomic ensemble of pseudo-spin-$1/2$
atoms following the proposal in~\cite{Christensen2013}. It can be
described in four simple steps. (i)~All spins are oriented in one
direction, and the prepared ensemble state is
$\ket{\Psi_0}=\ket{\uparrow\uparrow\ldots\uparrow\uparrow}$.  (ii)~A
single spin is probabilistically flipped into the opposite state,
without resolving which atom was affected such that the excitation is
distributed over the ensemble. The system state becomes
\begin{align}
  \ket{\Psi_1} & \equiv \frac{1}{\sqrt{\Natom}} \sum
  \limits_{l=1}^{\Natom} \ket{\uparrow\uparrow\ldots \uparrow
    \smash{\underset{\makebox[2pt]{\footnotesize$\overbrace{l\text{th
              atom}}$}}{\downarrow}}\uparrow \ldots \uparrow\uparrow}.
\end{align}
(iii))~A $\pi/2$ pulse acting homogeneously on all atoms transfers
each atom into an equal superposition: $\ket{\uparrow} \to \ket{+}$
and $\ket{\downarrow} \to \ket{-}$ with $\ket{\pm} \equiv
\frac{\ket{\uparrow} \pm \ket{\downarrow}} { \sqrt{2} } $.  Depending
on the presence or absence of the spin flip, this leaves the system in
one of the two following states:
\begin{align}
  \label{eq:rotatedstates}
  \ket{\Psi_0'} & = \ket{++\ldots++}, \\
  \ket{\Psi_1'} & = \frac{1}{\sqrt{\Natom}} \sum
  \limits_{l=1}^{\Natom}
  \ket{++\ldots+\smash{\underset{\makebox[2pt]{\footnotesize$\overbrace{l\text{th
              atom}}$}}{-}}+\ldots ++}.
\end{align}
(iv)~Atomic state analysis is performed by measuring the population
difference of atoms in the two spin states $\Delta N = N_{\uparrow} -
N_{\downarrow}$.

An interesting and somewhat counterintuitive observation is that the
probability distribution of the measurement outcome is fundamentally
altered whenever a single spin-flip has taken place. The magnitude of
the difference between the spin-flip and no-spin-flip distributions is
comparable to the atomic quantum projection noise ($\sim
\sqrt{\Natom}$) and is thus much bigger than an incoherent single-atom
effect. This enhancement is explained by quantum interference between
the single excited spin and the unaffected atoms.

\begin{figure}
  \centerline{
    \includegraphics[keepaspectratio,width=\columnwidth]{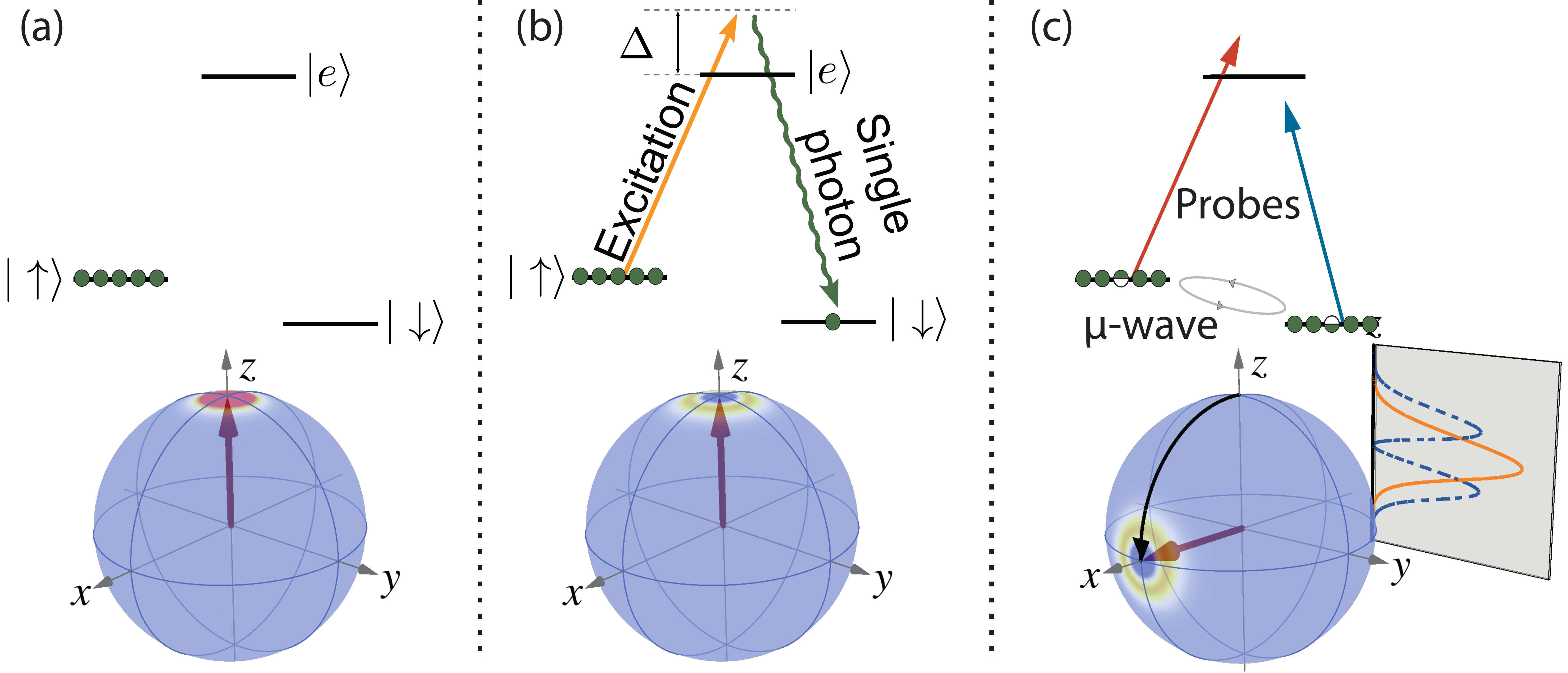}}
  \caption{\label{fig:creation}(Color online) The simplified atomic
    level structure and the collective Bloch sphere at different
    stages of the experiment. (a) All atoms are prepared in the
    $\ket{\uparrow}$ state via optical pumping. (b) Detection of a
    scattered anti-Stokes photon following a weak excitation of the
    ensemble signals that a single atom has been transferred to the
    $\ket{\downarrow}$ state. (c) A microwave $\pi/2$ pulse causes the
    single excited atom in~$\ket{\downarrow}$ to interfere with the
    remaining atoms in~$\ket{\uparrow}$ by rotating all spins into the
    equatorial plane. This creates the collective state
    $\ket{\Psi_1'}$, which is characterized by a continuous-variable
    measurement. The inset shows the probability density of
    $\hat{J}_z$ for $\ket{\Psi_0 '}$ [solid orange] and $\ket{\Psi_1
      '}$ [dashed blue].}
\end{figure}

The single-excitation state $\ket{\Psi_1'}$ has several interesting
features: in the limit of a large ensemble, $\Natom \gg 1$, the states
$\ket{\Psi_0'}$ and $\ket{\Psi_1'}$ correspond to the atomic
equivalent of the vacuum and single-excitation states of a bosonic
mode (Holstein-Primakoff approximation~\cite{Holstein1940}).  Unlike a
single-photon state which is superposed with a strong local oscillator
on a beam splitter to reveal its Wigner function~\cite{Lvovsky2009},
in the present case the atomic ensemble plays the role of the local
oscillator and is inseparable from the single spin carrying the
excitation.  The state $\ket{\Psi_1'}$ is non-Gaussian with a negative
Wigner function stored within a quantum memory. As such it is
potentially valuable for quantum information
applications~\cite{Ohliger2012, Kot2012, Duan2001}. This negativity of
the Wigner function leads to a non-Gaussian marginal distribution with
an increased variance compared to $\ket{\Psi_0'}$ [see inset in
\fref[c]{fig:creation}~\cite{Christensen2013}].  It is exactly this
increase that we will use to distinguish between the two states of
interest.  In our experiment various technical imperfections (detector
dark counts and so on.)  limit the purity of the $\ket{\Psi_1'}$-state
preparation. As shown in detail later, this reduces the expected
increase in the variance of the population difference. Due to the
signal enhancement by interference, even for a low-purity state we are
able to discriminate the created state against $\ket{\Psi_0'}$.

We employ an ensemble of approximately $\num{e5}$ cesium atoms; the
pseudospin system is formed by two stable levels in the ground-state
hyperfine manifold, the clock states $\ket{\uparrow} \equiv
\ket{F\!=\!4, m_F\!=\!0}$ and $\ket{\downarrow} \equiv \ket{F\!=\!3,
  m_F\!=\!0}$. Each atom $l$ is described by pseudospin operators
$2\hat
j_z^{(l)}=\ketbra{\uparrow}{\uparrow}^{(l)}-\ketbra{\downarrow}{\downarrow}^{(l)}$,
$2\hat j_x^{(l)}=\ketbra{+}{+}^{(l)}-\ketbra{-}{-}^{(l)}$, and
\mbox{$\hat{j}_y^{(l)} = i [\hat{j}_x^{(l)},\hat{j}_z^{(l)}]$.}
Introducing the collective operators $\hat{J}_i = \sum _{l=1}
^{\Natom} \hat{j}_i ^{(l)}$, the ensemble state can be visualized on
the Bloch sphere [see \fref{fig:creation}~\cite{Dowling1994}]. For the
characterization of the interference effect, the observable of
interest is $\hat{J}_z = \Delta N/2$.

\section{Experiment} \label{sec:experiment}

To create the state $\ket{\Psi_0}$, we first load atoms in a
magneto-optical trap (MOT), transfer them into a dipole trap (formed
by a $P \approx \SI{4.7}{\watt}$, $\SI{1064}{nm}$-laser beam), and
optically pump them into the $\ket{\downarrow}$ state using the $D_2$
line~\cite{Louchet-Chauvet2010}. With a microwave $\pi$ pulse and a
subsequent resonant $F=3 \rightarrow F'=4$ optical purifying pulse we
bring the atoms into the $\ket{\uparrow}$ state and remove any
remaining coherences between $\ket{\uparrow}$ and $\bra{\downarrow}$
[see \fref[a]{fig:creation}].  To minimize the inhomogeneous
broadening of the optical transitions, we briefly turn off the dipole
trap and subject the ensemble to a $\SI{2.5}{\micro \second}$
off-resonant excitation pulse, detuned by $\Delta=
\SI{5.4}{\mega\hertz}$ with respect to the $\ket{\uparrow}
\leftrightarrow \ket{e}=\ket{F'=4,m_{F'}=1}$ transition, focused to a
waist of $\SI{30}{\micro\meter}$ and comprising $n_{\text{exct}}
=\num{8.9e5}$ photons.  By independently measuring the reduction of
the microwave $\pi$ pulse contrast due to this excitation pulse, we
infer that $1-\eta_\text{scatter}=\SI{23}{\percent}$ of the atoms
scatter a photon from this excitation beam (see
\nameref{sec:scattering}); with a probability
$p_\text{forward}=\SI{1.43}{\percent}$ the atoms forward scatter a
photon with an energy corresponding to a decay to the
$\ket{\downarrow}$ state exactly into the detection spatial mode.  The
detection of a \emph{single} $\ket{\uparrow} \rightarrow F'=4
\rightarrow \ket{\downarrow}$ anti-Stokes photon signals that a single
atom has been transferred to the $\ket{\downarrow}$ state and thus
heralds the preparation of the $\ket{\Psi_1}$ state
[see~\fref[b]{fig:creation}~\cite{Duan2001}].  Using a microwave
$\pi/2$ pulse, we cause the single excited atom to interfere with the
remaining $\ket{\uparrow}$-state atoms [see \fref[c]{fig:creation}]
and reestablish the dipole trap.

\begin{figure}
  \centerline{\includegraphics[keepaspectratio,
    width=\columnwidth]{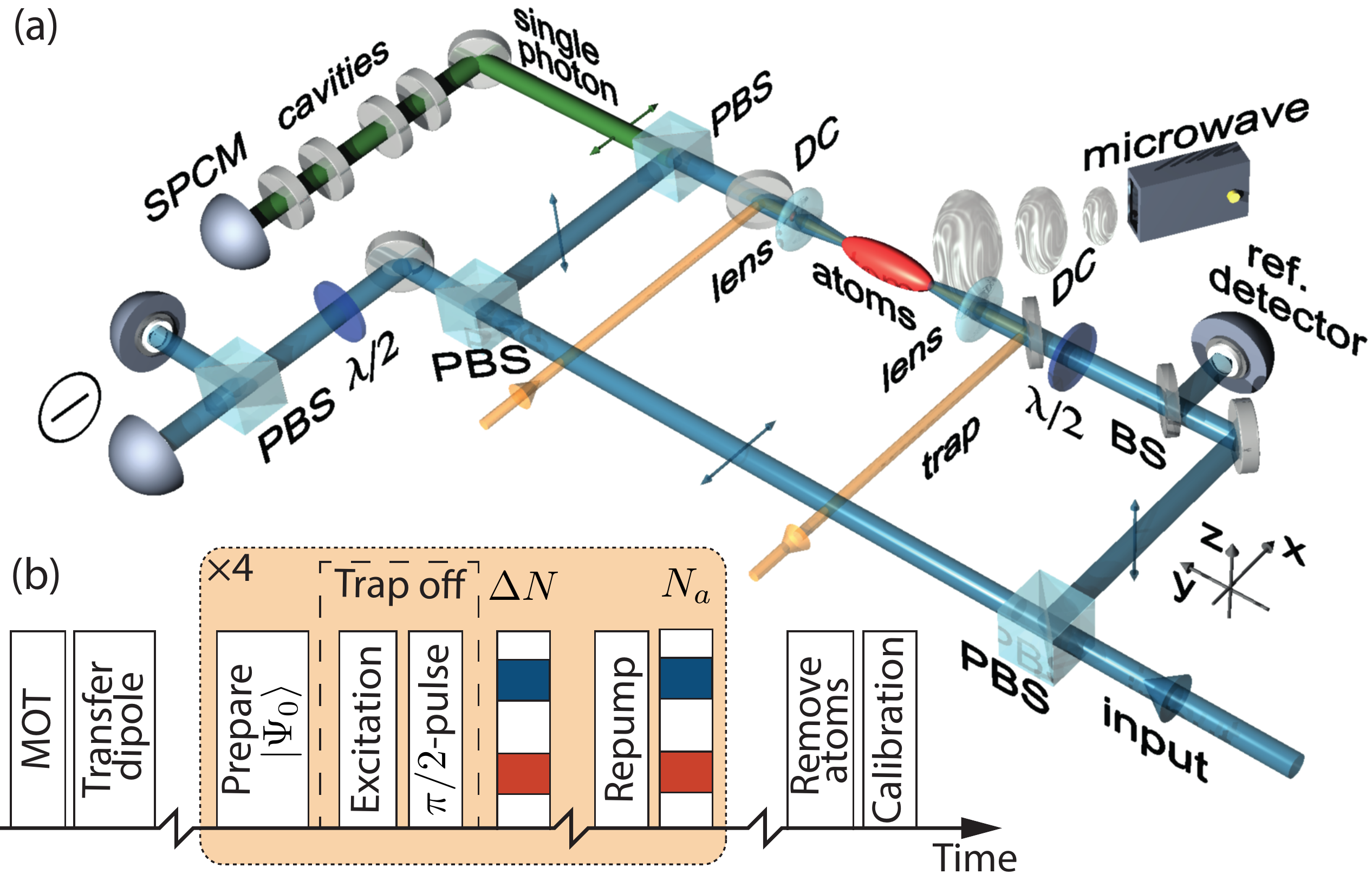}}
  \caption{\label{fig:setup}(Color online) (a) Experimental setup. The
    dipole-trapped atomic ensemble is overlapped with one arm of a
    MZI, using dichroic (DC) mirrors. The input mode of the MZI is
    used for the weak Raman excitation and for the dual-color QND
    measurement of atoms. A single-photon counter module (SPCM)
    detects the heralding photon. To select a photon in the desired
    decay channel, polarization (via PBS) and frequency (via
    Fabry–Pérot cavities) filters are implemented.  The atomic state
    is characterized by a dispersive QND measurement using balanced
    homodyne detection. A beam splitter (BS) is used to calibrate the
    probe power.  (b)~Pulse sequence.}
\end{figure}

The atoms are held in one arm of a Mach-Zehnder interferometer (MZI)
[see \fref[a]{fig:setup}], which allows us to measure $\hat J_z\propto
\Delta N$ by dual-color QND tomography~\cite{Louchet-Chauvet2010} with
a precision much better than the projection noise using
$n_{\text{probe}} = \num{1.51e8}$ photons in total [see
\fref[c]{fig:creation}]. We then repump all atoms into $F=4$ and
determine $\Natom$, by again measuring the atomic induced phase shift
\cite{Appel2009}.  Depending on the detection of the heralding
anti-Stokes photon, the measurement outcomes are associated with
$\hat{J}_z$ statistics of either the $\ket{\Psi_0'}$ or
$\ket{\Psi_1'}$ states.  To optimize the measurement time we reuse the
same MOT cloud four times, allowing us to measure the atomic state for
varying atom numbers.  Finally, the atoms are removed from the trap
using resonant light, and calibration measurements are performed [see
\fref[b]{fig:setup}]. To obtain the required statistics the experiment
is repeated more than a hundred thousand times.  The atomic tomography
method described above is an atomic analog of homodyne detection of
optical fields~\cite{Lvovsky2009}: the strong local oscillator field
is represented by the large number of atoms in the $\ket{\uparrow}$
state, and the quantum field is formed by the single
$\ket{\downarrow}$ state atom. The \num{50}:\num{50} beam splitter is
realized by the $\pi/2$ microwave pulse, and the intensity-difference
measurement is implemented by the measurement of $\Delta N$.

In order to enhance the probability of forward scattering a photon in
the desired channel $\ket{\uparrow} \rightarrow F'=4 \rightarrow
\ket{\downarrow}$ a bias magnetic field of $B = \SI{20.5}{\gauss}$ in
the $z$ direction [see \fref[a]{fig:setup}] is applied
~\cite{Christensen2013}.  Polarization and frequency filtering in the
heralding photon path (see \fref[a]{fig:setup}) is used to
discriminate the unwanted decay channels originating from $F'=4$. A
polarizing beam splitter cube (PBS), attenuating $\pi$-polarized light
by $\num{1/7e3}$, suppresses anti-Stokes photons leading to $\ket{F=3,
  4,m_F=\pm 1}$.  Two consecutive Fabry–Pérot cavities with a finesse
of $\mathcal{F} = \num{300}$ and a linewidth of $\delta \nu_c
=\SI{26}{\mega\hertz}$ filter out photons corresponding to decays into
$F=4$ states with a contrast of $\num{1/2.7e8}$. The decays into
$\ket{F=3, m_F = \pm 2}$ cannot be filtered out and present a
limitation on the purity of the state \cite{Christensen2013}.

\section{Analysis} \label{sec:analysis}

\begin{figure}
  \centerline{\includegraphics[keepaspectratio,
    width=\columnwidth]{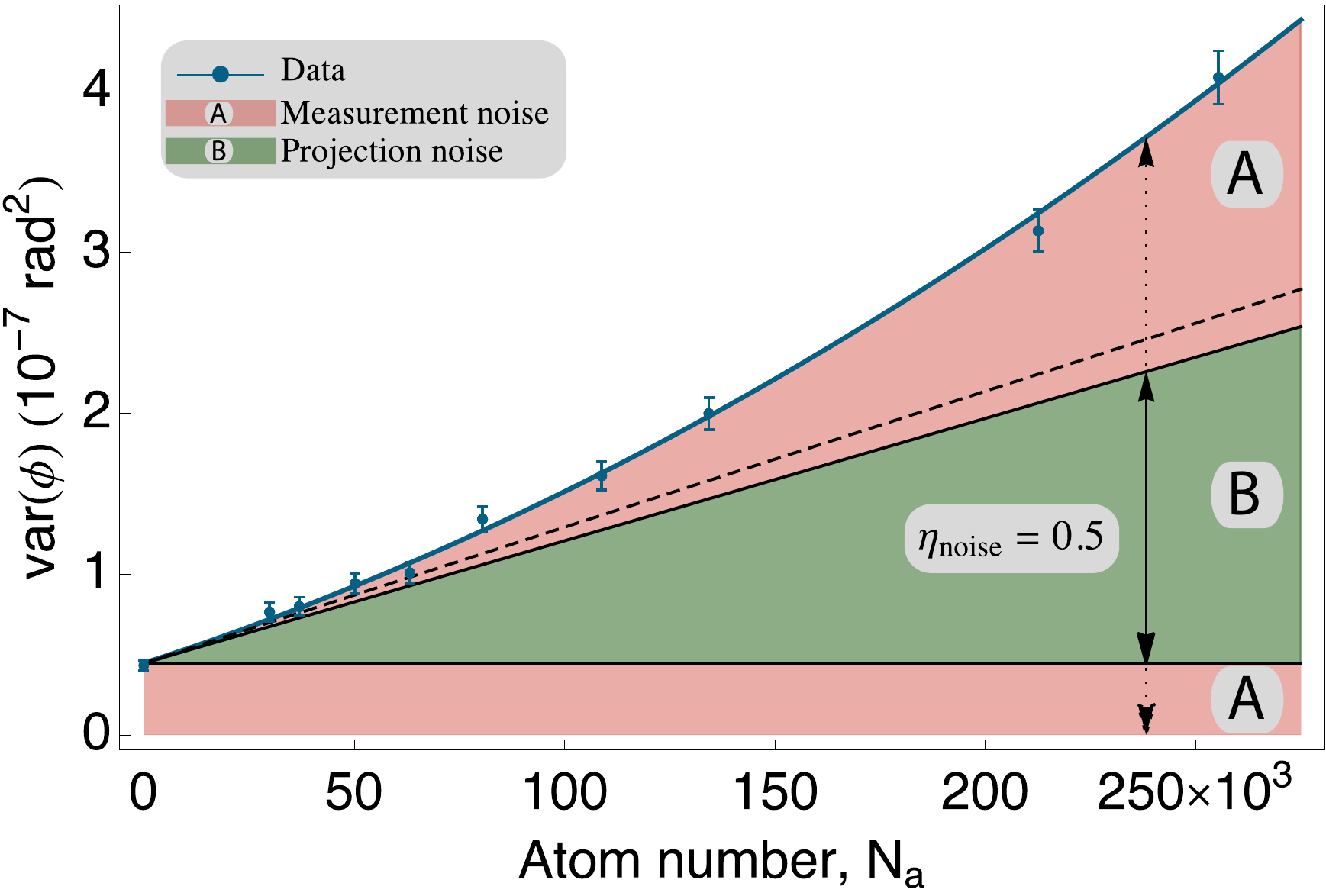}}
  \caption{\label{fig:noisescaling}(Color online) Variance of the
    measured optical phase shift $\phi$ as a function of the atom
    number. Different noise contributions are distinguished by a
    scaling analysis. The dominating linear part (dashed line)
    corresponds to atomic quantum projection noise; the quadratic and
    constant contributions (A) originate from technical fluctuations
    and light shot noise. Nine percent of the projection noise
    originates from noise-canceling reference measurements; the
    remaining fraction (B) constitutes
    $\eta_\text{noise}=\SI{50}{\percent}$ of the total noise.}
\end{figure}

The variable $\hat J_z\propto \Delta N$ is determined from the
differential phase shift $\tilde{\phi}_i$ imprinted by the atoms onto
two collinear laser beams of different frequency in a MZI
\cite{Louchet-Chauvet2010}.  The first step of the state analysis is
to calibrate the optical phase fluctuations to the atomic projection
noise.  Each measurement of $\tilde{\phi}_i$ is referenced to an
optimally weighted average of 12 measurements on an ensemble in
$\ket{\Psi_0 '}$ in order to reduce the effect of slow drifts in probe
powers.  A small additional amount (\SI{9}{\percent}) of light shot
noise and atomic projection noise from these reference measurements
causes a slight decrease of the detection efficiency
[see~\fref{fig:noisescaling} and~\nameref{sec:NoiseCancellation}].

A noise scaling analysis~\cite{Appel2009} confirms a predominant
linear scaling of the atomic noise with $N_a$, which is characteristic
for the atomic projection noise [see \fref{fig:noisescaling}].  This
linear scaling with $N_a$ is analogous to linear noise scaling with
the local oscillator power in photonic homodyne measurements.  For
larger $N_a$, we observe classical noise with its quadratic scaling.
The main contribution to this noise comes from frequency fluctuations
of the excitation pulse which cause classical fluctuations in the atom
number difference between the $F=3$ and $F=4$ hyperfine manifolds.
Additionally, in the bias magnetic field the
$\ket{\uparrow}\leftrightarrow\ket{\downarrow}$ transition frequency
becomes sensitive to magnetic fields
($\SI{17.5}{\kilo\hertz/\gauss}$), such that magnetic-field
fluctuations can affect the quality of the microwave pulses.

\begin{figure}
  \centerline{\includegraphics[keepaspectratio,
    width=\columnwidth]{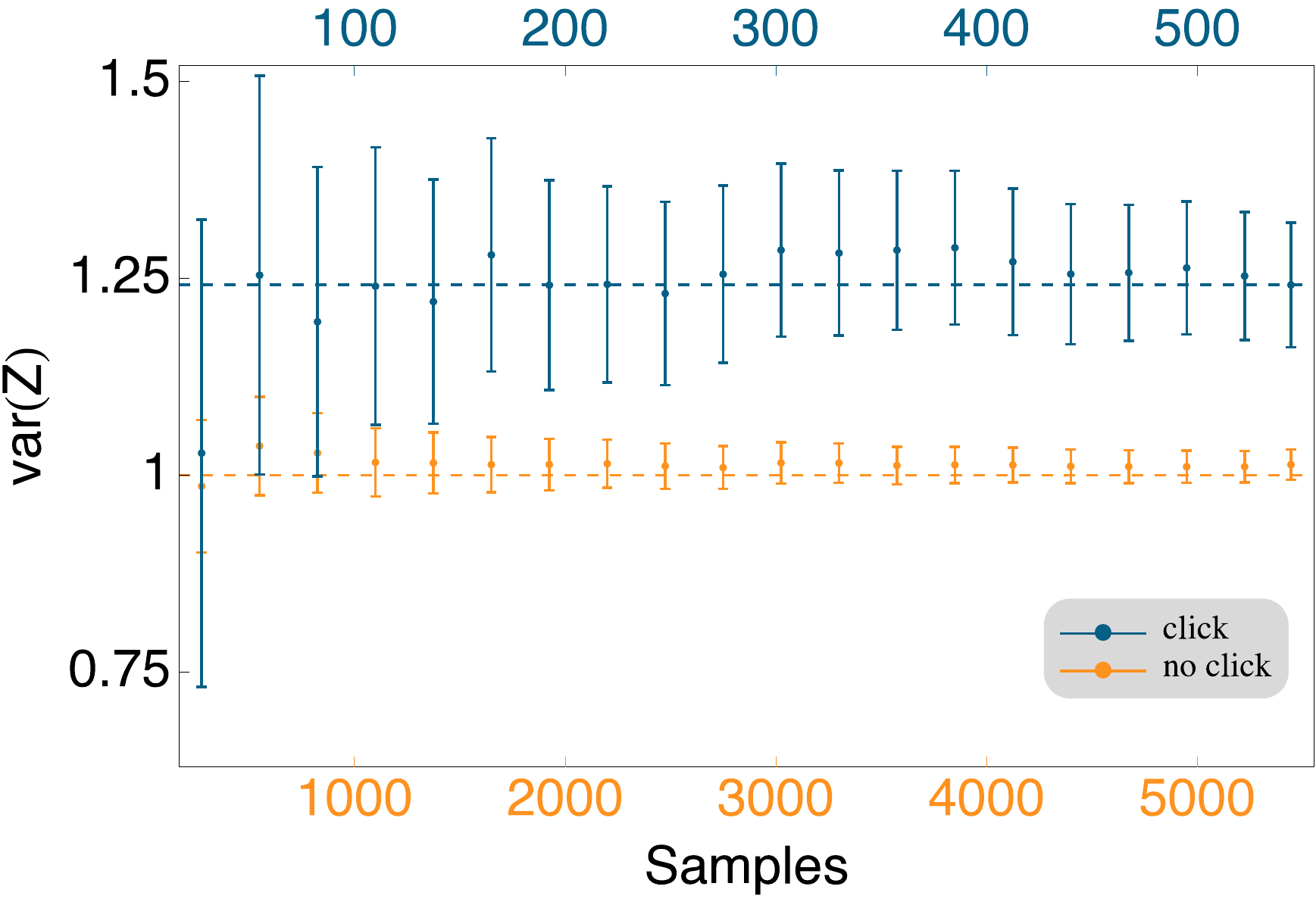}}
  \caption{\label{fig:cumstat}(Color online) Cumulative statistics for
    the variance of the measurement outcomes of $\Delta N$ for the two
    created states, showing an increased variance for heralding
    events. The sample variance is plotted against the number of
    observations with a heralding photon (no heralding photon) as
    depicted on the top (bottom) axis.  }
\end{figure}

The second part of our analysis addresses the comparison of the
$\hat{J}_z$ probability distributions of states $\ket{\Psi_0 '}$ and
$\ket{\Psi_1 '}$. Here we include only the data with $\Natom
>\num{2e5}$, where the probability of detecting the heralding photon
is the highest.  In order to compensate for slow drifts of the
light-atom coupling strength, we introduce a noise normalization
procedure and divide each $\Delta N$ measurement by the standard
deviation of the neighboring $M = \num{200}$ measurement
outcomes. This allows us to locally normalize the variance to unity
for events where \emph{no} heralding photon was detected.  The results
for the normalized variances $Z$ for the two states as a function of
the number of samples are presented in \fref{fig:cumstat}.  We find
\begin{align}
  \var(Z^{\text{no click}}) &= \num{1.02} \pm \num{0.02} \\
  \var(Z^{\text{click}})&= \num{1.24} \pm \num{0.08},
\end{align}
where the errors correspond to one standard deviation of the variance
estimator (see \nameref{sec:state-discrimination}).  As expected from
our normalization procedure, in the case of no heralding photon, we
obtain the unity variance.  In the case of the presence of the
heralding photon, a statistically significant increase of
\SI{24}{\percent} in the variance is observed.

Heralding errors convert the pure target state into a statistical
mixture described by a density operator
\begin{equation}
  \hat{\rho} = p \ket{\Psi_1'}\bra{\Psi_1'}
  + (1-p) \ket{\Psi_0'}\bra{\Psi_0'}.\label{eq:rho}
\end{equation}
Here $p$ is the classical probability that $\ket{\Psi_1'}$ is actually
prepared when a photon is detected. For state $\hat{\rho}$ we would
expect
\begin{align}
  \var(\Delta N)_{\hat \rho} & = p \bra{\Psi_1'} 4\hat J_z^2
  \ket{\Psi_1'} +  (1-p) \bra{\Psi_0'} 4\hat J_z^2 \ket{\Psi_0'} \nonumber \\
  & = 3 p \Natom + (1-p) \Natom.
  \label{eq:vardeltanexpect}
\end{align}
Heralding errors which reduce the purity of the state include the
detector dark counts with $p_\text{dark}=\num{0.13} p_\text{click}$,
leakage of the excitation pulse through the filters with
$p_\text{exct}=\num{0.38} p_\text{click}$, and unfiltered photons
originating from the decay into $\ket{F=3,m_F=\pm2}$ states with
$p_\text{decay}=\num{0.11}p_\text{click}$. Here
$p_\text{click}=\num{6.7e-3}$ is the observed photon counting
probability per excitation pulse. For the purity of the created state
we find
\begin{align}
  p_\text{state} & =
  1-\frac{p_\text{dark}+p_\text{decay}+p_\text{exct}}{p_\text{click}}=
  \SI{38}{\percent}. \label{eq:pExpected}
\end{align}
With a stronger suppression of the false-positive events by better
filtering cavities a state purity exceeding $\SI{70}{\percent}$ can be
foreseen.

The quantum efficiency of the atomic state detection is finite due to
several effects which add state-independent Gaussian noise.  It is
well known \cite{Appel2007} that when homodyne quadrature measurements
are normalized to a vacuum state with added uncorrelated Gaussian
noise, this effectively decreases the quantum efficiency
$\eta_\text{Q}$ of the detection. This decrease can be modeled by
assuming a vacuum admixture of $1-\eta_\text{Q}$ followed by an ideal
detection.  In our experiment such additional noise that is
uncorrelated with the quantum state of interest [red areas (dark gray)
in \fref{fig:noisescaling}] originates from the electronic detector
noise, photon shot noise ~\cite{Kiesel2012, Christensen2013},
classical fluctuations in the atomic state initialization, and the
noise from the 12 reference measurements
(\nameref{sec:NoiseCancellation}). These noise sources lead to an
effective detection quantum efficiency of
$\eta_\text{noise}=\SI{50}{\percent}$ as indicated in
\fref{fig:noisescaling}. The non-perfect overlap between the
excitation and the photon-collection modes contributes
$\eta_{\text{mm}} = \SI{75}{\percent}$.  Spontaneous emission of
photons into modes which do not interfere with the single excitation
acts similar to the imperfect spatial overlap with the local
oscillator in photonic homodyning and leads to the factor
$\eta_\text{scatter} = \SI{77}{\percent} $.  Finally, the phase
mismatch between the prepared spin wave and the detection mode caused
by the inhomogeneous ac-Stark shift induced by the excitation beam and
the refractive index of the atomic ensemble leads to a minor
correction of $\eta_{\text{inhom}} =\SI{92}{\percent}$ (see
\nameref{sec:phase-mismatch}).  The total efficiency of the state
detection is given by $\eta_Q = \eta_\text{noise} \eta_\text{mm}
\eta_\text{inhom} \eta_\text{scatter}=\SI{27}{\percent}$.  The
expected variance of the created state can then be found as
$\var(\Delta N)_{\hat \rho} /\Natom = 3 p_\text{expect} +
(1-p_\text{expect}) =1.20$, where $p_\text{expect} =p_\text{state}\,
\eta_\text{Q}$, in good agreement with the experimental value.

The contribution of multiple excitations of the spin-wave mode which,
in principle, can strongly affect the results \cite{Laurat2006,
  Kuzmich2003} is negligible in our experiment. This is a result of a
relatively high probability to detect photons scattered into the
detection mode of $p_{\text{d}} = \SI{18}{\percent}$ combined with a
photon-number-resolving detector (the detector dead time of
\SI{50}{ns} is short compared to the \SI{2.5}{\micro \second}
excitation pulse duration).  A detailed calculation
(\nameref{sec:multiple-spin-wave}) reveals that, on the condition of a
heralding photon, the probability to find more than one atomic
excitation is $p(n>1 | \text{1click}) = \num{9e-3}$ which increases
$\var(Z^\text{click})$ by $\SI{2}{\percent}$. The two-excitation
contribution amounts to less than $\SI{17}{\percent}$ of that of a
coherent spin state with the same mean excitation number.

\section{Conclusion} \label{sec:conclusion}

In conclusion, we have implemented a hybrid discrete-continuous
protocol where a collective single spin excitation heralded by the
detection of a single photon is characterized by a direct measurement
of a collective continuous-variable atomic operator. Although, in
general, an observed increase in the variance of the atomic operator
could be due to classical reasons, such as an admixture of a thermal
state, in our experiment the increase is solely due to the detection
of a scattered single photon. Even stronger evidence of the successful
generation of a single excitation state requires determination of
higher-order statistical moments, which in turn demands a higher
purity of the produced quantum state
\cite{Dubost2012,Schmied2011,Lvovsky2009}.  Steps towards this goal
could include a stronger light-atom coupling achievable in ensembles
trapped around nanofibers~\cite{Vetsch2010, Lacroute2012}, atoms
coupled to optical resonators~\cite{Leroux2010a, Zhang2012,
  Colombe2007}, or photonic structures~\cite{Thompson2013,
  Goban2013}. Furthermore, a better suppression of false photon counts
and using atoms with a simpler level structure (e.g. $ {}^{87}$Rb)
could help. These improvements of the method should allow for
observation of a negative Wigner function of a macroscopic atomic
ensemble, certifying its non-classical properties~\cite{Kot2012}.
Such a state is a building block for atomic Schrödinger's cat
states~\cite{Brask2010}; it can be used in precision
measurements~\cite{McConnell2013,Simon2011} and provides a
non-Gaussian resource for future quantum information
processing~\cite{Ohliger2012}.

\begin{acknowledgments}
  This work is funded by the Danish National Research Foundation, EU
  projects SIQS, MALICIA, ERC grant INTERFACE, and DARPA through the
  project QuASAR. We thank Emil Zeuthen and Anders S. Sørensen for
  helpful discussions.
\end{acknowledgments}

\appendix

\section{Technical fluctuations} \label{sec:NoiseCancellation}

Using the dual-color QND measurement we prepare and probe our
ensemble, obtaining measurement outcomes $\tilde{\phi}_i$. To
eliminate technical fluctuations, we subtract the baseline of the
empty interferometer. Further noise reduction is achieved by
performing 12 reference measurements $\{\varphi_i^{j}\}_{j \in
  \{-6,\ldots -1,1,\ldots 6\}}$ on a $\ket{\Psi_0'}$ state, six each
immediately before and after $\tilde{\phi}_i$ is measured. Since these
measurements are performed on independently prepared atomic ensembles,
all correlations between them are of technical nature. We therefore
decorrelate $\tilde \phi_i$ from its reference measurements
$\varphi_i^j$ by subtracting the correlated noise contributions:
\begin{align}
  \phi_i = \tilde{\phi_i} - \sum \limits_{\substack{j=-6 \\ j\neq
      0}}^{6} w_j \, \varphi_i^j.
\end{align}
The 12 weight factors $w_j$ are chosen such that the sample variance
$\var(\{\phi_i\})$ is minimized.  Since each reference measurement
contains both the full (uncorrelated) atomic projection and shot
noise, this procedure not only reduces technical fluctuations but also
adds $\sum_j w_j^2=\num{0.09}$ units of projection and shot noise to
each $\phi_i$ measurement. This decreases the state detection
efficiency $\eta_\text{noise}$, as explained in \sref{sec:analysis}.
The above choice of $w_j$ guarantees an optimization of this
trade-off.

\section{State discrimination} \label{sec:state-discrimination}

To compare the measurement statistics for the two created states
$\ket{\Psi_1'} $ and $\ket{\Psi_0'}$, we only consider data with
$\Natom > \num{2e5}$. For these high-atom-number realizations we
obtain a high $\eta_{\text{noise}}$, which directly leads to a large
increase in the difference of variances, as explained in
\sref{sec:analysis}.

As our data are acquired over a duration of 2 weeks, we observe slow,
long-term changes in the variance of the measurements.  These are
caused mainly by drifts in the relative optical power of the MOT
beams, changes in the background vapor pressure due to operation of
the cesium dispensers, and accumulation of dust particles in the
shared optical path of the strong, focused dipole trap, excitation,
and probe beams.  We therefore perform a local noise normalization to
avoid long-term drifts in the variance of our measurements.

For each correlation-removed measurement outcome $\phi_i$ we compute
the sample variance of the $M$ surrounding experiments:
\begin{align}
  Y_i \equiv \var \left( \{ \phi_{i-M/2}, \ldots, \phi_{i+M/2}\}
  \right),
\end{align}
and we use this to normalize the variance of $\phi_i$ to the
surrounding data points:
\begin{align}
  Z_i \equiv \frac{\phi_i}{\sqrt{Y_i}}.
\end{align}
The $Z_i$ originating from a $\ket{\Psi_0'}$ measurement, by this
construction, have an average variance of $\approx 1$ since
$p_\text{click}\ll 1$; that is $Y_i$ contain almost entirely no-click
events.

Our final parameter of interest is the sample variance of a set of
$Z_i$
\begin{align}
  W_{\mathbf{L}} \equiv \var \left( \{ Z_{i}\}_{i \in \mathbf{L}}
  \right),
\end{align}
both for the sets of indices $\mathbf{L^\text{click}} = \{ i :
\text{click detected} \}$ and $\mathbf{L^\text{no click}} = \{ i :
\text{no click} \}$. In the following we focus on estimating the
statistical uncertainty $\delta W_{\mathbf{L}}$ on $W_{\mathbf{L}}$
and denote with $L=|\mathbf{L}|$ the number of samples used in the
calculation.

Due to the finite number of points used in estimating $Y_i$ there is a
statistical uncertainty on this estimator which carries over to
$Z_i$. Since all $\phi_i$ from within the range $[i-M/2,\ldots,i+M/2]$
can be considered independently and identically distributed, we can
give the uncertainty of $Y_i$ simply by the mean-square error (MSE) of
the variance estimator and find
\begin{align}
  \delta Y_i = \sqrt{\frac{2}{M-1}} Y_i.
\end{align}

This allows us to find the variance of each of the $Z_i$ by Taylor
expansion around $\langle Y_i\rangle$ as
\begin{align}
  \var(Z_i) = \left(1+ \frac{1}{4} \frac{2}{M+1} \right) \var \left(
    \frac{\phi_i}{\langle Y_i \rangle} \right).\label{eq:varZi}
\end{align}

One complication is that within a range of $M$ neighboring experiments
the $Z_i$ are not statistically independent any longer due to our
normalization procedure. As can be seen from~\eqref{eq:varZi}, by
ensuring that $M \gg 1$, we can make the contribution of correlated
noise to each $Z_i$ negligibly small. If, additionally, we either
choose $L \gg M$ or ensure that the members of $\mathbf{L}$ are spaced
much farther than $M$ on average, $W_\mathbf{L}$ is an unbiased
estimator of the $Z_i$ variance, and its uncertainty is
\begin{align}
  \delta W_\mathbf{L} = \sqrt{\frac{2}{L-1}} W_\mathbf{L},
\end{align}
which is simply the MSE on the variance estimator for $L$ independent
samples.

We confirm all our error estimates experimentally both by dividing our
data set into subsets and evaluating the standard deviation of the
variance estimates and by the bootstrapping method (resampling).

\section{Scattering} \label{sec:scattering}

\begin{figure}
  \centerline{\includegraphics[keepaspectratio,
    width=0.5\textwidth]{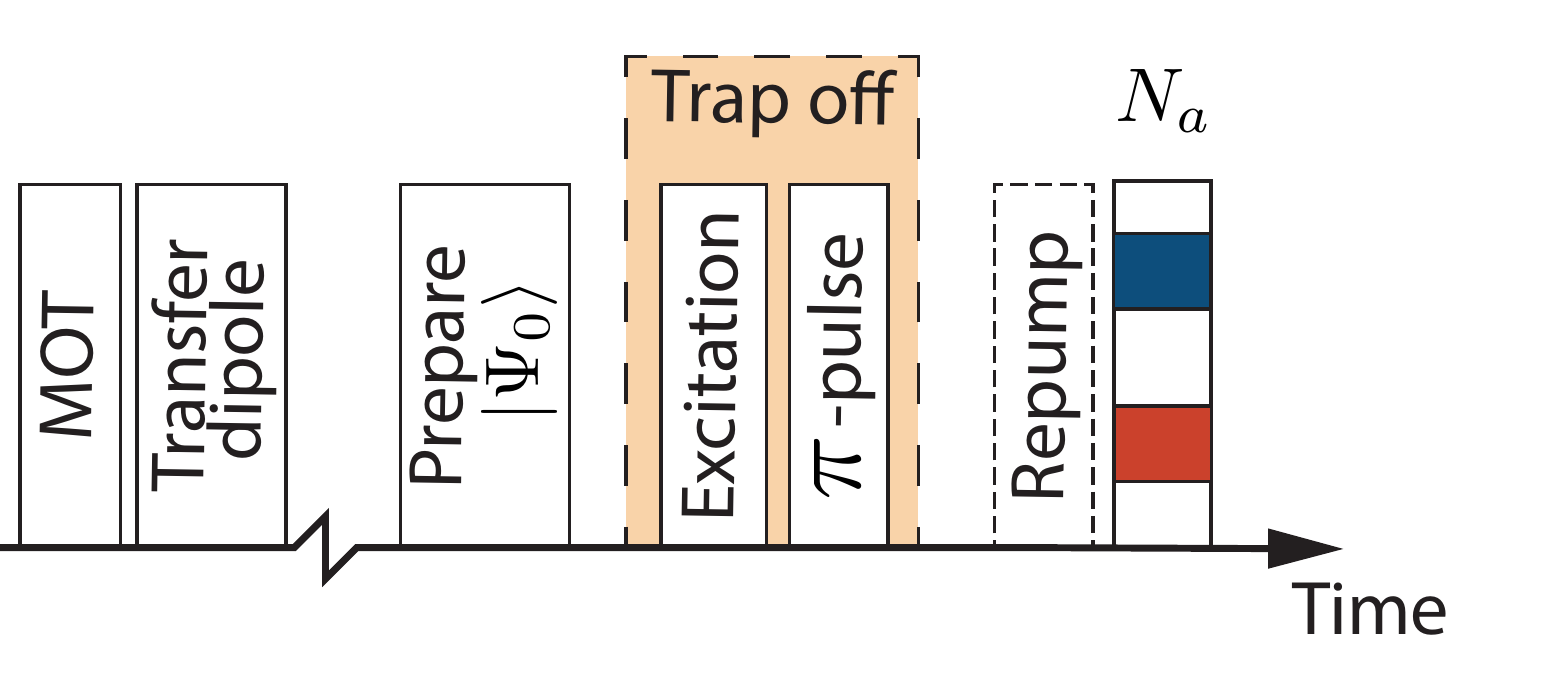}}
  \caption{(Color online) Experimental sequence in order to estimate
    $\eta_{\text{scatter}}$.}
  \label{fig:scattering}
\end{figure}

To determine $\eta_{\text{scatter}}$, the fraction of atoms that
scatter a photon from the excitation pulse, we perform a separate
calibration experiment [see \fref{fig:scattering}].  First we prepare
all atoms in the $\ket{\uparrow}$ state (see \sref{sec:experiment}).
While the trap is switched off we send the excitation pulse followed
by a microwave $\pi$ pulse and an optional repumping pulse, resonant
to the $F=3 \to F'=4$ transition. Finally the number of atoms $\Natom$
in $F=4$ is determined.  In the presence of the repumping pulse, all
atoms in the trap are detected, whereas in the absence of the
repumping pulse only atoms scattered into the $\ket{F=4,m_F\neq 0}$
and $\ket{F=3,m_F = 0}$ states are measured.  Finally, using the
relevant Clebsch-Gordan coefficients to determine the fraction of
atoms scattering into each Zeeman state, we can find the fraction of
atoms that undergo a scattering event to be
$1-\eta_\text{scatter}=\SI{23}{\percent}$.

\section{Phase mismatch}\label{sec:phase-mismatch}

Inhomogeneous phase shifts can reduce the interference visibility
$\eta_{\text{inhom}}=\eta_\text{phase} \eta_\text{ac-Stark}$. Here we
consider two effects.

\subsection{Longitudinal phase profile}\label{sec:long-phase-prof}

The first effect concerns the refractive index mismatch between the
scattered single photon and the excitation beam. In our
one-dimensional model we describe this by a position dependent phase
difference $\theta (y) = \Delta k \, y$, where $\Delta k =
k_{\mu\text{-wave}} + k_\text{exct} - k_\text{photon}$ is the
wave-vector mismatch between microwave-, excitation- and
heralding-photon fields. Since no atoms reside in $\ket{\downarrow}$,
the atomic phase mismatch emerges exclusively from the optical phase
shift of the excitation beam $\chi_\text{exct} = \theta(\La)$.

For the sake of simplicity we assume a homogeneous atomic density
distribution and average the phase over the length of the atomic
ensemble $\La$:
\begin{align}
  \eta_{\text{phase}}= \left | \frac{1}{\La} \int_0 ^{\La} \e^{-i
      \theta(y)} \dd y \right|^2 = \sinc^2( \chi_\text{exct}/2 )
\end{align}

We can relate the excitation beam phase shift $\chi_\text{exct}$ to
the measured phase shift $\chi_\text{probe}$ of the QND probe during
the atom number measurement. Let $\alpha_0$ denote the resonant
optical depth on a closed transition. Then, light traveling through a
medium with optical depth $\alpha_0$, detuned by a frequency
$\Delta_i$ with respect to a transition with relative strength $\wp_i$
will experience an optical phase shift,
\begin{align}
  \chi(\Delta,\alpha_0) = \frac{\alpha_0}{4} \sum \limits_i \wp_i
  \frac{\Delta_i}{\Delta_i^2 + (\Gamma/2)^2}.
\end{align}

When all our atoms are pumped into $F'=4$, with the probe beam we
measure an optical phase shift corresponding to $\alpha_0=31$, from
which, using the Clebsch-Gordan coefficients and the detuning
corresponding to our excitation beam, we obtain $\chi_\text{exct}=
\SI{42}{\degree}$, which gives $\eta_{\text{phase}} =
\SI{95}{\percent}$.

\subsection{Transversal phase profile} \label{sec:transv-phase-prof}

The off-resonant excitation pulse leads to an ac-Stark shift, both of
the $\ket{\uparrow}$ state and the excited states. Only the spatially
inhomogeneous shift of the $\ket{\uparrow}$ state affects the spin
wave coherence, which reduces the interference visibility. Since the
longitudinal extent of our atomic ensemble is short compared to the
Rayleigh length of the light beams, we restrict our model to
transversal effects, which can be evaluated as \cite{Hammerer2010}
\begin{align}
  \eta_{\text{ac-Stark}} = \frac{\int \left | \iint \varrho(x,z) \,
      I(x,z )^2 \, \e^{-i \omega_{\text{LS}}(x,z) t}\, \dd x \dd z
    \right |^2 \dd t} {\tau \left | \iint \varrho(x,z) \, I(x,z)^2 \,
      \dd x \dd z\right |^2}
\end{align}
where $I(x,z)$ denotes the transverse Gaussian intensity profile of
the excitation beam, $\varrho(x,z)$ is the atomic column density,
$\omega_{\text{LS}}(x,y)\propto I(x,z)$ is the ac-Stark shift of the
$\ket{\uparrow}$ state, and $\tau$ is the excitation pulse
duration. Numerically evaluating the above allows us to estimate the
effect and we find $\eta_{\text{ac-Stark}} = \SI{97}{\percent}$.

\section{Multiple spin-wave
  excitations} \label{sec:multiple-spin-wave}

When the single photon counter reports a ``click'', this can originate
from dark counts, leakage of excitation photons, or actual Stokes
photons scattered from the atomic ensemble (either from the desired or
from other unwanted transitions). To investigate the influence of
multiple-Stokes-photon events on our analysis, we calculate $p \left
  (n | 1\text{click} \right)$. This is the probability that the atomic
ensemble scattered $n$ Stokes photons with an energy corresponding to
a decay to the state $\ket{\downarrow}$ exactly into the detection
spatial mode, on the condition of detecting a single click.  Since the
detector dead time of \SI{50}{\nano \second} is much smaller than the
\SI{2.5}{\micro \second} excitation pulse length, we effectively have
a number-resolving photon detection. By Bayes' rule we have
\begin{align}
  p(n | 1\text{click})= \frac{ p (n) \, p(1\text{click} | n)
  }{p(1\text{click}) },\label{eq:pn1click}
\end{align}
where $p(n)$ is the probability to scatter $n$ photons with an energy
corresponding to a decay to the state $\ket{\downarrow}$ exactly into
the detection spatial mode and $p(\text{1click})$ is the probability
to detect exactly one click.  For a two-mode squeezed vacuum state the
photon number statistics in the individual modes is thermal
\cite{Lee1990}. Thus the probability to find $n$ Stokes photons
corresponding to a decay into $\ket{\downarrow}$ is given by
\begin{align}
  p_{S0}(n) = \left(1-p_{\text{0}}\right){p_{\text{0}}}^n,
\end{align}
where $p_\text{0}$ is the probability to generate at least one of the
desired Stokes photons.

Dark counts and leakage photons from the excitation pulse follow a
Poisson distribution
\begin{align}
  p_\text{DE}(n) = \frac{ {p_\text{f}}^n \, e^{-p_\text{f}}}{n!},
\end{align}
where $p_\text{f}$ is the mean number of such false-positive clicks in
the absence of atoms.  Stokes photons corresponding to decay into
$\ket{F=3,m_F=\pm 2}$ are not filtered out and therefore also cause
false positives. Their generation is distributed as
\begin{align}
  p_\text{S2}(n)=(1-p_\text{2}){p_\text{2}}^n,
\end{align}
where $p_\text{2}$ is the probability to scatter at least one photon
corresponding to a decay to $\ket{F=3,m_F=\pm 2}$.

Finally, we introduce $p_\text{d}=0.8^2 \times 0.56 \times 0.5$, the
probability that a single Stokes photon in the detection mode causes
an (additional) click which is given by the product of the
transmission coefficients through the filter cavities, through other
optics, and by the detector quantum efficiency. The complementary
probability is denoted $\tilde{p_\text{d}}=1-p_\text{d}$.

Then the probability for detecting $n$ additional clicks due to
unwanted $\ket{F=3,m_F=\pm2}$ Stokes photons is
\begin{align}
  p_\text{DS2}(n) & = \sum \limits_{k=n}^\infty p_\text{S2}(k) \,
  \binom{k}{k-n} \,  \tilde{p}_\text{d}{}^{k-n} \, {p_\text{d}}^n \\
  & = \frac{ (1-p_\text{2}) \, ( p_\text{2} \, p_\text{d} )^n } {
    \left( 1- p_\text{2} \, \tilde{p}_\text{d} \right)^{n+1}}.
\end{align}

The probability to find no false-positive events is $p_\text{F}(0)=
p_\text{DE}(0) \, p_\text{DS2}(0)$, whereas the probability to find
exactly one false positive is $p_\text{F}(1)= p_\text{DE}(1) \,
p_\text{DS2}(0) + p_\text{DE}(0) \, p_\text{DS2}(1)$.

With this, $p( \text{1click}|n )$, the probability to detect exactly
one click when the atoms scatter $n$ photons corresponding to the
desired $\ket{\downarrow}$ decay, is made up of two cases: (a) exactly
one of the $n$ photons makes it through the filtering optical elements
and causes a click in the detector while simultaneously no
false-positive counts are detected, and (b) none of the $n$ photons
cause a detector click while exactly one false positive count is
detected:
\begin{align}
  p \left ( 1\text{click}|n\right) & = n \, p_d \,
  \tilde{p}_\text{d}{}^{n-1} \, p_\text{F}(0) +
  {\tilde{p}_\text{d}}{}^n \, p_\text{F}(1).
\end{align}

Using the relation $\sum _{n=0}^\infty p(n|\text{1click})=1$, from
\eqref{eq:pn1click} we can determine $p( \text{1click})$ and obtain:

\begin{align}
  p(n|1\text{click}) & = \frac{ \tilde{p}_\text{d}{}^n \,
    {p_\text{0}}{}^n (1- \tilde{p}_\text{{d}} \, p_\text{0}) \left(n
      \frac{p_\text{d}}{ \tilde{p}_\text{{d}}} +p_\text{f} + \frac{p_d
        \, p_\text{2}}{1-{\tilde{p}_\text{{d}}}\, p_\text{2}}\right)}
  { p_\text{f} + \frac{p_\text{d}\,
      p_\text{2}}{1-{\tilde{p}_{\text{d}}} \, p_\text{2}} +
    \frac{p_\text{d} \, p_\text{0}}{1-{\tilde{p}_\text{{d}}} \,
      p_\text{0}}}.
\end{align}

Since we know $p_\text{f}=p_\text{dark} + p_\text{exct}$ from
reference measurements without atoms and $p_\text{2}= 0.3\,
p_\text{0}$ from the ratio of the transition strengths, we can deduce
$p_\text{0}=p_\text{forward} = 0.014$ and obtain the probability
$p(n|1\text{click})$ for different $n$:
\begin{align}
  p\left(n=0|1\text{click}\right) &= 0.606 \\
  p\left(n=1|1\text{click}\right) &= 0.385 \\
  p\left(n=2|1\text{click}\right) &= 0.009
\end{align}

Since all multi-excitation contributions together only amount to
$\SI{2.4}{\percent}$ of the single excitation contribution, we can
safely neglect their influence in our analysis.

\bibliographystyle{apsrev4-1} 
\bibliography{SingleExcitation}

\begin{thebibliography}{41}%
\makeatletter
\providecommand \@ifxundefined [1]{%
 \@ifx{#1\undefined}
}%
\providecommand \@ifnum [1]{%
 \ifnum #1\expandafter \@firstoftwo
 \else \expandafter \@secondoftwo
 \fi
}%
\providecommand \@ifx [1]{%
 \ifx #1\expandafter \@firstoftwo
 \else \expandafter \@secondoftwo
 \fi
}%
\providecommand \natexlab [1]{#1}%
\providecommand \enquote  [1]{``#1''}%
\providecommand \bibnamefont  [1]{#1}%
\providecommand \bibfnamefont [1]{#1}%
\providecommand \citenamefont [1]{#1}%
\providecommand \href@noop [0]{\@secondoftwo}%
\providecommand \href [0]{\begingroup \@sanitize@url \@href}%
\providecommand \@href[1]{\@@startlink{#1}\@@href}%
\providecommand \@@href[1]{\endgroup#1\@@endlink}%
\providecommand \@sanitize@url [0]{\catcode `\\12\catcode `\$12\catcode
  `\&12\catcode `\#12\catcode `\^12\catcode `\_12\catcode `\%12\relax}%
\providecommand \@@startlink[1]{}%
\providecommand \@@endlink[0]{}%
\providecommand \url  [0]{\begingroup\@sanitize@url \@url }%
\providecommand \@url [1]{\endgroup\@href {#1}{\urlprefix }}%
\providecommand \urlprefix  [0]{URL }%
\providecommand \Eprint [0]{\href }%
\providecommand \doibase [0]{http://dx.doi.org/}%
\providecommand \selectlanguage [0]{\@gobble}%
\providecommand \bibinfo  [0]{\@secondoftwo}%
\providecommand \bibfield  [0]{\@secondoftwo}%
\providecommand \translation [1]{[#1]}%
\providecommand \BibitemOpen [0]{}%
\providecommand \bibitemStop [0]{}%
\providecommand \bibitemNoStop [0]{.\EOS\space}%
\providecommand \EOS [0]{\spacefactor3000\relax}%
\providecommand \BibitemShut  [1]{\csname bibitem#1\endcsname}%
\let\auto@bib@innerbib\@empty
\bibitem [{\citenamefont {Hammerer}\ \emph {et~al.}(2010)\citenamefont
  {Hammerer}, \citenamefont {Sørensen},\ and\ \citenamefont
  {Polzik}}]{Hammerer2010}%
  \BibitemOpen
  \bibfield  {author} {\bibinfo {author} {\bibfnamefont {K.}~\bibnamefont
  {Hammerer}}, \bibinfo {author} {\bibfnamefont {A.}~\bibnamefont {Sørensen}},
  \ and\ \bibinfo {author} {\bibfnamefont {E.}~\bibnamefont {Polzik}},\ }\href
  {\doibase 10.1103/RevModPhys.82.1041} {\bibfield  {journal} {\bibinfo
  {journal} {Rev. Mod. Phys.}\ }\textbf {\bibinfo {volume} {82}},\ \bibinfo
  {pages} {1041} (\bibinfo {year} {2010})}\BibitemShut {NoStop}%
\bibitem [{\citenamefont {Kimble}(2008)}]{Kimble2008}%
  \BibitemOpen
  \bibfield  {author} {\bibinfo {author} {\bibfnamefont {H.~J.}\ \bibnamefont
  {Kimble}},\ }\href {\doibase 10.1038/nature07127} {\bibfield  {journal}
  {\bibinfo  {journal} {Nature}\ }\textbf {\bibinfo {volume} {453}},\ \bibinfo
  {pages} {1023} (\bibinfo {year} {2008})}\BibitemShut {NoStop}%
\bibitem [{\citenamefont {Choi}\ \emph {et~al.}(2010)\citenamefont {Choi},
  \citenamefont {Goban}, \citenamefont {Papp}, \citenamefont {van Enk},\ and\
  \citenamefont {Kimble}}]{Choi2010}%
  \BibitemOpen
  \bibfield  {author} {\bibinfo {author} {\bibfnamefont {K.~S.}\ \bibnamefont
  {Choi}}, \bibinfo {author} {\bibfnamefont {A.}~\bibnamefont {Goban}},
  \bibinfo {author} {\bibfnamefont {S.~B.}\ \bibnamefont {Papp}}, \bibinfo
  {author} {\bibfnamefont {S.~J.}\ \bibnamefont {van Enk}}, \ and\ \bibinfo
  {author} {\bibfnamefont {H.~J.}\ \bibnamefont {Kimble}},\ }\href {\doibase
  10.1038/nature09568} {\bibfield  {journal} {\bibinfo  {journal} {Nature}\
  }\textbf {\bibinfo {volume} {468}},\ \bibinfo {pages} {412} (\bibinfo {year}
  {2010})}\BibitemShut {NoStop}%
\bibitem [{\citenamefont {MacRae}\ \emph {et~al.}(2012)\citenamefont {MacRae},
  \citenamefont {Brannan}, \citenamefont {Achal},\ and\ \citenamefont
  {Lvovsky}}]{MacRae2012}%
  \BibitemOpen
  \bibfield  {author} {\bibinfo {author} {\bibfnamefont {A.}~\bibnamefont
  {MacRae}}, \bibinfo {author} {\bibfnamefont {T.}~\bibnamefont {Brannan}},
  \bibinfo {author} {\bibfnamefont {R.}~\bibnamefont {Achal}}, \ and\ \bibinfo
  {author} {\bibfnamefont {A.~I.}\ \bibnamefont {Lvovsky}},\ }\href {\doibase
  10.1103/PhysRevLett.109.033601} {\bibfield  {journal} {\bibinfo  {journal}
  {Phys. Rev. Lett.}\ }\textbf {\bibinfo {volume} {109}},\ \bibinfo {pages}
  {033601} (\bibinfo {year} {2012})}\BibitemShut {NoStop}%
\bibitem [{\citenamefont {Duan}\ \emph {et~al.}(2001)\citenamefont {Duan},
  \citenamefont {Lukin}, \citenamefont {Cirac},\ and\ \citenamefont
  {Zoller}}]{Duan2001}%
  \BibitemOpen
  \bibfield  {author} {\bibinfo {author} {\bibfnamefont {L.~M.}\ \bibnamefont
  {Duan}}, \bibinfo {author} {\bibfnamefont {M.~D.}\ \bibnamefont {Lukin}},
  \bibinfo {author} {\bibfnamefont {J.~I.}\ \bibnamefont {Cirac}}, \ and\
  \bibinfo {author} {\bibfnamefont {P.}~\bibnamefont {Zoller}},\ }\href
  {\doibase 10.1038/35106500} {\bibfield  {journal} {\bibinfo  {journal}
  {Nature}\ }\textbf {\bibinfo {volume} {414}},\ \bibinfo {pages} {413}
  (\bibinfo {year} {2001})}\BibitemShut {NoStop}%
\bibitem [{\citenamefont {Specht}\ \emph {et~al.}(2011)\citenamefont {Specht},
  \citenamefont {Nölleke}, \citenamefont {Reiserer}, \citenamefont {Uphoff},
  \citenamefont {Figueroa}, \citenamefont {Ritter},\ and\ \citenamefont
  {Rempe}}]{Specht2011}%
  \BibitemOpen
  \bibfield  {author} {\bibinfo {author} {\bibfnamefont {H.~P.}\ \bibnamefont
  {Specht}}, \bibinfo {author} {\bibfnamefont {C.}~\bibnamefont {Nölleke}},
  \bibinfo {author} {\bibfnamefont {A.}~\bibnamefont {Reiserer}}, \bibinfo
  {author} {\bibfnamefont {M.}~\bibnamefont {Uphoff}}, \bibinfo {author}
  {\bibfnamefont {E.}~\bibnamefont {Figueroa}}, \bibinfo {author}
  {\bibfnamefont {S.}~\bibnamefont {Ritter}}, \ and\ \bibinfo {author}
  {\bibfnamefont {G.}~\bibnamefont {Rempe}},\ }\href {\doibase
  10.1038/nature09997} {\bibfield  {journal} {\bibinfo  {journal} {Nature}\
  }\textbf {\bibinfo {volume} {473}},\ \bibinfo {pages} {190} (\bibinfo {year}
  {2011})}\BibitemShut {NoStop}%
\bibitem [{\citenamefont {Krauter}\ \emph {et~al.}(2013)\citenamefont
  {Krauter}, \citenamefont {Salart}, \citenamefont {Muschik}, \citenamefont
  {Petersen}, \citenamefont {Shen}, \citenamefont {Fernholz},\ and\
  \citenamefont {Polzik}}]{Krauter2013}%
  \BibitemOpen
  \bibfield  {author} {\bibinfo {author} {\bibfnamefont {H.}~\bibnamefont
  {Krauter}}, \bibinfo {author} {\bibfnamefont {D.}~\bibnamefont {Salart}},
  \bibinfo {author} {\bibfnamefont {C.~A.}\ \bibnamefont {Muschik}}, \bibinfo
  {author} {\bibfnamefont {J.~M.}\ \bibnamefont {Petersen}}, \bibinfo {author}
  {\bibfnamefont {H.}~\bibnamefont {Shen}}, \bibinfo {author} {\bibfnamefont
  {T.}~\bibnamefont {Fernholz}}, \ and\ \bibinfo {author} {\bibfnamefont
  {E.~S.}\ \bibnamefont {Polzik}},\ }\href {\doibase 10.1038/nphys2631}
  {\bibfield  {journal} {\bibinfo  {journal} {Nature Physics}\ }\textbf
  {\bibinfo {volume} {9}},\ \bibinfo {pages} {400} (\bibinfo {year}
  {2013})}\BibitemShut {NoStop}%
\bibitem [{\citenamefont {Appel}\ \emph {et~al.}(2009)\citenamefont {Appel},
  \citenamefont {Windpassinger}, \citenamefont {Oblak}, \citenamefont
  {Busk~Hoff}, \citenamefont {Kjærgaard},\ and\ \citenamefont
  {Polzik}}]{Appel2009}%
  \BibitemOpen
  \bibfield  {author} {\bibinfo {author} {\bibfnamefont {J.}~\bibnamefont
  {Appel}}, \bibinfo {author} {\bibfnamefont {P.~J.}\ \bibnamefont
  {Windpassinger}}, \bibinfo {author} {\bibfnamefont {D.}~\bibnamefont
  {Oblak}}, \bibinfo {author} {\bibfnamefont {U.}~\bibnamefont {Busk~Hoff}},
  \bibinfo {author} {\bibfnamefont {N.}~\bibnamefont {Kjærgaard}}, \ and\
  \bibinfo {author} {\bibfnamefont {E.~S.}\ \bibnamefont {Polzik}},\ }\href
  {\doibase 10.1073/pnas.0901550106} {\bibfield  {journal} {\bibinfo  {journal}
  {P. Natl. Acad. Sci.}\ }\textbf {\bibinfo {volume} {106}},\ \bibinfo {pages}
  {10960} (\bibinfo {year} {2009})}\BibitemShut {NoStop}%
\bibitem [{\citenamefont {Leroux}\ \emph {et~al.}(2010)\citenamefont {Leroux},
  \citenamefont {Schleier-Smith},\ and\ \citenamefont
  {Vuletić}}]{Leroux2010a}%
  \BibitemOpen
  \bibfield  {author} {\bibinfo {author} {\bibfnamefont {I.~D.}\ \bibnamefont
  {Leroux}}, \bibinfo {author} {\bibfnamefont {M.~H.}\ \bibnamefont
  {Schleier-Smith}}, \ and\ \bibinfo {author} {\bibfnamefont {V.}~\bibnamefont
  {Vuletić}},\ }\href {\doibase 10.1103/PhysRevLett.104.073602} {\bibfield
  {journal} {\bibinfo  {journal} {Phys. Rev. Lett.}\ }\textbf {\bibinfo
  {volume} {104}},\ \bibinfo {eid} {073602} (\bibinfo {year}
  {2010})}\BibitemShut {NoStop}%
\bibitem [{\citenamefont {Fernholz}\ \emph {et~al.}(2008)\citenamefont
  {Fernholz}, \citenamefont {Krauter}, \citenamefont {Jensen}, \citenamefont
  {Sherson}, \citenamefont {Sørensen},\ and\ \citenamefont
  {Polzik}}]{Fernholz2008}%
  \BibitemOpen
  \bibfield  {author} {\bibinfo {author} {\bibfnamefont {T.}~\bibnamefont
  {Fernholz}}, \bibinfo {author} {\bibfnamefont {H.}~\bibnamefont {Krauter}},
  \bibinfo {author} {\bibfnamefont {K.}~\bibnamefont {Jensen}}, \bibinfo
  {author} {\bibfnamefont {J.~F.}\ \bibnamefont {Sherson}}, \bibinfo {author}
  {\bibfnamefont {A.~S.}\ \bibnamefont {Sørensen}}, \ and\ \bibinfo {author}
  {\bibfnamefont {E.~S.}\ \bibnamefont {Polzik}},\ }\href {\doibase
  10.1103/PhysRevLett.101.073601} {\bibfield  {journal} {\bibinfo  {journal}
  {Phys. Rev. Lett.}\ }\textbf {\bibinfo {volume} {101}},\ \bibinfo {eid}
  {073601} (\bibinfo {year} {2008})}\BibitemShut {NoStop}%
\bibitem [{\citenamefont {Wasilewski}\ \emph {et~al.}(2010)\citenamefont
  {Wasilewski}, \citenamefont {Jensen}, \citenamefont {Krauter}, \citenamefont
  {Renema}, \citenamefont {Balabas},\ and\ \citenamefont
  {Polzik}}]{Wasilewski2010}%
  \BibitemOpen
  \bibfield  {author} {\bibinfo {author} {\bibfnamefont {W.}~\bibnamefont
  {Wasilewski}}, \bibinfo {author} {\bibfnamefont {K.}~\bibnamefont {Jensen}},
  \bibinfo {author} {\bibfnamefont {H.}~\bibnamefont {Krauter}}, \bibinfo
  {author} {\bibfnamefont {J.~J.}\ \bibnamefont {Renema}}, \bibinfo {author}
  {\bibfnamefont {M.~V.}\ \bibnamefont {Balabas}}, \ and\ \bibinfo {author}
  {\bibfnamefont {E.~S.}\ \bibnamefont {Polzik}},\ }\href {\doibase
  10.1103/PhysRevLett.104.133601} {\bibfield  {journal} {\bibinfo  {journal}
  {Phys. Rev. Lett.}\ }\textbf {\bibinfo {volume} {104}},\ \bibinfo {pages}
  {133601} (\bibinfo {year} {2010})}\BibitemShut {NoStop}%
\bibitem [{\citenamefont {Louchet-Chauvet}\ \emph {et~al.}(2010)\citenamefont
  {Louchet-Chauvet}, \citenamefont {Appel}, \citenamefont {Renema},
  \citenamefont {Oblak}, \citenamefont {Kjærgaard},\ and\ \citenamefont
  {Polzik}}]{Louchet-Chauvet2010}%
  \BibitemOpen
  \bibfield  {author} {\bibinfo {author} {\bibfnamefont {A.}~\bibnamefont
  {Louchet-Chauvet}}, \bibinfo {author} {\bibfnamefont {J.}~\bibnamefont
  {Appel}}, \bibinfo {author} {\bibfnamefont {J.~J.}\ \bibnamefont {Renema}},
  \bibinfo {author} {\bibfnamefont {D.}~\bibnamefont {Oblak}}, \bibinfo
  {author} {\bibfnamefont {N.}~\bibnamefont {Kjærgaard}}, \ and\ \bibinfo
  {author} {\bibfnamefont {E.~S.}\ \bibnamefont {Polzik}},\ }\href {\doibase
  10.1088/1367-2630/12/6/065032} {\bibfield  {journal} {\bibinfo  {journal}
  {New. J. Phys.}\ }\textbf {\bibinfo {volume} {12}},\ \bibinfo {pages}
  {065032} (\bibinfo {year} {2010})}\BibitemShut {NoStop}%
\bibitem [{\citenamefont {Koschorreck}\ \emph {et~al.}(2010)\citenamefont
  {Koschorreck}, \citenamefont {Napolitano}, \citenamefont {Dubost},\ and\
  \citenamefont {Mitchell}}]{Koschorreck2010}%
  \BibitemOpen
  \bibfield  {author} {\bibinfo {author} {\bibfnamefont {M.}~\bibnamefont
  {Koschorreck}}, \bibinfo {author} {\bibfnamefont {M.}~\bibnamefont
  {Napolitano}}, \bibinfo {author} {\bibfnamefont {B.}~\bibnamefont {Dubost}},
  \ and\ \bibinfo {author} {\bibfnamefont {M.~W.}\ \bibnamefont {Mitchell}},\
  }\href {\doibase 10.1103/PhysRevLett.104.093602} {\bibfield  {journal}
  {\bibinfo  {journal} {Phys. Rev. Lett.}\ }\textbf {\bibinfo {volume} {104}},\
  \bibinfo {eid} {093602} (\bibinfo {year} {2010})}\BibitemShut {NoStop}%
\bibitem [{\citenamefont {Simon}\ \emph {et~al.}(2010)\citenamefont {Simon},
  \citenamefont {Afzelius}, \citenamefont {Appel}, \citenamefont {Boyer~de
  La~Giroday}, \citenamefont {Dewhurst}, \citenamefont {Gisin}, \citenamefont
  {Hu}, \citenamefont {Jelezko}, \citenamefont {Kröll}, \citenamefont
  {Müller}, \citenamefont {Nunn}, \citenamefont {Polzik}, \citenamefont
  {Rarity}, \citenamefont {de~Riedmatten}, \citenamefont {Rosenfeld},
  \citenamefont {Shields}, \citenamefont {Sköld}, \citenamefont {Stevenson},
  \citenamefont {Thew}, \citenamefont {Walmsley}, \citenamefont {Weber},
  \citenamefont {Weinfurter}, \citenamefont {Wrachtrup},\ and\ \citenamefont
  {Young}}]{Simon2010}%
  \BibitemOpen
  \bibfield  {author} {\bibinfo {author} {\bibfnamefont {C.}~\bibnamefont
  {Simon}}, \bibinfo {author} {\bibfnamefont {M.}~\bibnamefont {Afzelius}},
  \bibinfo {author} {\bibfnamefont {J.}~\bibnamefont {Appel}}, \bibinfo
  {author} {\bibfnamefont {A.}~\bibnamefont {Boyer~de La~Giroday}}, \bibinfo
  {author} {\bibfnamefont {S.~J.}\ \bibnamefont {Dewhurst}}, \bibinfo {author}
  {\bibfnamefont {N.}~\bibnamefont {Gisin}}, \bibinfo {author} {\bibfnamefont
  {C.~Y.}\ \bibnamefont {Hu}}, \bibinfo {author} {\bibfnamefont
  {F.}~\bibnamefont {Jelezko}}, \bibinfo {author} {\bibfnamefont
  {S.}~\bibnamefont {Kröll}}, \bibinfo {author} {\bibfnamefont {J.~H.}\
  \bibnamefont {Müller}}, \bibinfo {author} {\bibfnamefont {J.}~\bibnamefont
  {Nunn}}, \bibinfo {author} {\bibfnamefont {E.~S.}\ \bibnamefont {Polzik}},
  \bibinfo {author} {\bibfnamefont {J.~G.}\ \bibnamefont {Rarity}}, \bibinfo
  {author} {\bibfnamefont {H.}~\bibnamefont {de~Riedmatten}}, \bibinfo {author}
  {\bibfnamefont {W.}~\bibnamefont {Rosenfeld}}, \bibinfo {author}
  {\bibfnamefont {A.~J.}\ \bibnamefont {Shields}}, \bibinfo {author}
  {\bibfnamefont {N.}~\bibnamefont {Sköld}}, \bibinfo {author} {\bibfnamefont
  {R.~M.}\ \bibnamefont {Stevenson}}, \bibinfo {author} {\bibfnamefont
  {R.}~\bibnamefont {Thew}}, \bibinfo {author} {\bibfnamefont {I.~A.}\
  \bibnamefont {Walmsley}}, \bibinfo {author} {\bibfnamefont {M.~C.}\
  \bibnamefont {Weber}}, \bibinfo {author} {\bibfnamefont {H.}~\bibnamefont
  {Weinfurter}}, \bibinfo {author} {\bibfnamefont {J.}~\bibnamefont
  {Wrachtrup}}, \ and\ \bibinfo {author} {\bibfnamefont {R.~J.}\ \bibnamefont
  {Young}},\ }\href {\doibase 10.1140/epjd/e2010-00103-y} {\bibfield  {journal}
  {\bibinfo  {journal} {Eur. Phys. J. D}\ }\textbf {\bibinfo {volume} {58}},\
  \bibinfo {pages} {1} (\bibinfo {year} {2010})}\BibitemShut {NoStop}%
\bibitem [{\citenamefont {Neergaard-Nielsen}\ \emph {et~al.}(2006)\citenamefont
  {Neergaard-Nielsen}, \citenamefont {Nielsen}, \citenamefont {Hettich},
  \citenamefont {Mølmer},\ and\ \citenamefont
  {Polzik}}]{Neergaard-Nielsen2006}%
  \BibitemOpen
  \bibfield  {author} {\bibinfo {author} {\bibfnamefont {J.~S.}\ \bibnamefont
  {Neergaard-Nielsen}}, \bibinfo {author} {\bibfnamefont {B.~M.}\ \bibnamefont
  {Nielsen}}, \bibinfo {author} {\bibfnamefont {C.}~\bibnamefont {Hettich}},
  \bibinfo {author} {\bibfnamefont {K.}~\bibnamefont {Mølmer}}, \ and\
  \bibinfo {author} {\bibfnamefont {E.~S.}\ \bibnamefont {Polzik}},\ }\href
  {\doibase 10.1103/PhysRevLett.97.083604} {\bibfield  {journal} {\bibinfo
  {journal} {Phys. Rev. Lett.}\ }\textbf {\bibinfo {volume} {97}},\ \bibinfo
  {eid} {083604} (\bibinfo {year} {2006})}\BibitemShut {NoStop}%
\bibitem [{\citenamefont {Ourjoumtsev}\ \emph {et~al.}(2007)\citenamefont
  {Ourjoumtsev}, \citenamefont {Jeong}, \citenamefont {Tualle-Brouri},\ and\
  \citenamefont {Grangier}}]{Ourjoumtsev2007}%
  \BibitemOpen
  \bibfield  {author} {\bibinfo {author} {\bibfnamefont {A.}~\bibnamefont
  {Ourjoumtsev}}, \bibinfo {author} {\bibfnamefont {H.}~\bibnamefont {Jeong}},
  \bibinfo {author} {\bibfnamefont {R.}~\bibnamefont {Tualle-Brouri}}, \ and\
  \bibinfo {author} {\bibfnamefont {P.}~\bibnamefont {Grangier}},\ }\href
  {\doibase 10.1038/nature06054} {\bibfield  {journal} {\bibinfo  {journal}
  {Nature}\ }\textbf {\bibinfo {volume} {448}},\ \bibinfo {pages} {784}
  (\bibinfo {year} {2007})}\BibitemShut {NoStop}%
\bibitem [{\citenamefont {Takeda}\ \emph {et~al.}(2013)\citenamefont {Takeda},
  \citenamefont {Mizuta}, \citenamefont {Fuwa}, \citenamefont {van Loock},\
  and\ \citenamefont {Furusawa}}]{Takeda2013}%
  \BibitemOpen
  \bibfield  {author} {\bibinfo {author} {\bibfnamefont {S.}~\bibnamefont
  {Takeda}}, \bibinfo {author} {\bibfnamefont {T.}~\bibnamefont {Mizuta}},
  \bibinfo {author} {\bibfnamefont {M.}~\bibnamefont {Fuwa}}, \bibinfo {author}
  {\bibfnamefont {P.}~\bibnamefont {van Loock}}, \ and\ \bibinfo {author}
  {\bibfnamefont {A.}~\bibnamefont {Furusawa}},\ }\href {\doibase
  10.1038/nature12366} {\bibfield  {journal} {\bibinfo  {journal} {Nature}\
  }\textbf {\bibinfo {volume} {500}},\ \bibinfo {pages} {315} (\bibinfo {year}
  {2013})}\BibitemShut {NoStop}%
\bibitem [{\citenamefont {Lvovsky}\ \emph {et~al.}(2013)\citenamefont
  {Lvovsky}, \citenamefont {Ghobadi}, \citenamefont {Chandra}, \citenamefont
  {Prasad},\ and\ \citenamefont {Simon}}]{Lvovsky2013}%
  \BibitemOpen
  \bibfield  {author} {\bibinfo {author} {\bibfnamefont {A.~I.}\ \bibnamefont
  {Lvovsky}}, \bibinfo {author} {\bibfnamefont {R.}~\bibnamefont {Ghobadi}},
  \bibinfo {author} {\bibfnamefont {A.}~\bibnamefont {Chandra}}, \bibinfo
  {author} {\bibfnamefont {A.~S.}\ \bibnamefont {Prasad}}, \ and\ \bibinfo
  {author} {\bibfnamefont {C.}~\bibnamefont {Simon}},\ }\href {\doibase
  10.1038/nphys2682} {\bibfield  {journal} {\bibinfo  {journal} {Nature
  Physics}\ }\textbf {\bibinfo {volume} {9}},\ \bibinfo {pages} {541} (\bibinfo
  {year} {2013})}\BibitemShut {NoStop}%
\bibitem [{\citenamefont {Bruno}\ \emph {et~al.}(2013)\citenamefont {Bruno},
  \citenamefont {Martin}, \citenamefont {Sekatski}, \citenamefont {Sangouard},
  \citenamefont {Thew},\ and\ \citenamefont {Gisin}}]{Bruno2013}%
  \BibitemOpen
  \bibfield  {author} {\bibinfo {author} {\bibfnamefont {N.}~\bibnamefont
  {Bruno}}, \bibinfo {author} {\bibfnamefont {A.}~\bibnamefont {Martin}},
  \bibinfo {author} {\bibfnamefont {P.}~\bibnamefont {Sekatski}}, \bibinfo
  {author} {\bibfnamefont {N.}~\bibnamefont {Sangouard}}, \bibinfo {author}
  {\bibfnamefont {R.~T.}\ \bibnamefont {Thew}}, \ and\ \bibinfo {author}
  {\bibfnamefont {N.}~\bibnamefont {Gisin}},\ }\href {\doibase
  10.1038/nphys2681} {\bibfield  {journal} {\bibinfo  {journal} {Nature
  Physics}\ }\textbf {\bibinfo {volume} {9}},\ \bibinfo {pages} {545} (\bibinfo
  {year} {2013})}\BibitemShut {NoStop}%
\bibitem [{\citenamefont {Brask}\ \emph {et~al.}(2010)\citenamefont {Brask},
  \citenamefont {Rigas}, \citenamefont {Polzik}, \citenamefont {Andersen},\
  and\ \citenamefont {S{\o}rensen}}]{Brask2010}%
  \BibitemOpen
  \bibfield  {author} {\bibinfo {author} {\bibfnamefont {J.~B.}\ \bibnamefont
  {Brask}}, \bibinfo {author} {\bibfnamefont {I.}~\bibnamefont {Rigas}},
  \bibinfo {author} {\bibfnamefont {E.~S.}\ \bibnamefont {Polzik}}, \bibinfo
  {author} {\bibfnamefont {U.~L.}\ \bibnamefont {Andersen}}, \ and\ \bibinfo
  {author} {\bibfnamefont {A.~S.}\ \bibnamefont {S{\o}rensen}},\ }\href
  {\doibase 10.1103/PhysRevLett.105.160501} {\bibfield  {journal} {\bibinfo
  {journal} {Phys. Rev. Lett.}\ }\textbf {\bibinfo {volume} {105}},\ \bibinfo
  {pages} {160501} (\bibinfo {year} {2010})}\BibitemShut {NoStop}%
\bibitem [{\citenamefont {McConnell}\ \emph {et~al.}(2013)\citenamefont
  {McConnell}, \citenamefont {Zhang}, \citenamefont {{\'C}uk}, \citenamefont
  {Hu}, \citenamefont {Schleier-Smith},\ and\ \citenamefont
  {Vuleti{\'c}}}]{McConnell2013}%
  \BibitemOpen
  \bibfield  {author} {\bibinfo {author} {\bibfnamefont {R.}~\bibnamefont
  {McConnell}}, \bibinfo {author} {\bibfnamefont {H.}~\bibnamefont {Zhang}},
  \bibinfo {author} {\bibfnamefont {S.}~\bibnamefont {{\'C}uk}}, \bibinfo
  {author} {\bibfnamefont {J.}~\bibnamefont {Hu}}, \bibinfo {author}
  {\bibfnamefont {M.~H.}\ \bibnamefont {Schleier-Smith}}, \ and\ \bibinfo
  {author} {\bibfnamefont {V.}~\bibnamefont {Vuleti{\'c}}},\ }\href {\doibase
  10.1103/PhysRevA.88.063802} {\bibfield  {journal} {\bibinfo  {journal}
  {Physical Review A}\ }\textbf {\bibinfo {volume} {88}},\ \bibinfo {pages}
  {063802} (\bibinfo {year} {2013})}\BibitemShut {NoStop}%
\bibitem [{\citenamefont {Christensen}\ \emph {et~al.}(2013)\citenamefont
  {Christensen}, \citenamefont {Béguin}, \citenamefont {Sørensen},
  \citenamefont {Bookjans}, \citenamefont {Oblak}, \citenamefont {Müller},
  \citenamefont {Appel},\ and\ \citenamefont {Polzik}}]{Christensen2013}%
  \BibitemOpen
  \bibfield  {author} {\bibinfo {author} {\bibfnamefont {S.~L.}\ \bibnamefont
  {Christensen}}, \bibinfo {author} {\bibfnamefont {J.~B.}\ \bibnamefont
  {Béguin}}, \bibinfo {author} {\bibfnamefont {H.~L.}\ \bibnamefont
  {Sørensen}}, \bibinfo {author} {\bibfnamefont {E.}~\bibnamefont {Bookjans}},
  \bibinfo {author} {\bibfnamefont {D.}~\bibnamefont {Oblak}}, \bibinfo
  {author} {\bibfnamefont {J.~H.}\ \bibnamefont {Müller}}, \bibinfo {author}
  {\bibfnamefont {J.}~\bibnamefont {Appel}}, \ and\ \bibinfo {author}
  {\bibfnamefont {E.~S.}\ \bibnamefont {Polzik}},\ }\href {\doibase
  10.1088/1367-2630/15/1/015002} {\bibfield  {journal} {\bibinfo  {journal}
  {New J. Phys., Focus issue on Quantum Tomography}\ }\textbf {\bibinfo
  {volume} {15}},\ \bibinfo {pages} {015002} (\bibinfo {year}
  {2013})}\BibitemShut {NoStop}%
\bibitem [{\citenamefont {Holstein}\ and\ \citenamefont
  {Primakoff}(1940)}]{Holstein1940}%
  \BibitemOpen
  \bibfield  {author} {\bibinfo {author} {\bibfnamefont {T.}~\bibnamefont
  {Holstein}}\ and\ \bibinfo {author} {\bibfnamefont {H.}~\bibnamefont
  {Primakoff}},\ }\href {\doibase 10.1103/PhysRev.58.1098} {\bibfield
  {journal} {\bibinfo  {journal} {Phys. Rev.}\ }\textbf {\bibinfo {volume}
  {58}},\ \bibinfo {pages} {1098} (\bibinfo {year} {1940})}\BibitemShut
  {NoStop}%
\bibitem [{\citenamefont {Lvovsky}\ and\ \citenamefont
  {Raymer}(2009)}]{Lvovsky2009}%
  \BibitemOpen
  \bibfield  {author} {\bibinfo {author} {\bibfnamefont {A.~I.}\ \bibnamefont
  {Lvovsky}}\ and\ \bibinfo {author} {\bibfnamefont {M.~G.}\ \bibnamefont
  {Raymer}},\ }\href {\doibase 10.1103/RevModPhys.81.299} {\bibfield  {journal}
  {\bibinfo  {journal} {Rev. Mod. Phys.}\ }\textbf {\bibinfo {volume} {81}},\
  \bibinfo {pages} {299} (\bibinfo {year} {2009})}\BibitemShut {NoStop}%
\bibitem [{\citenamefont {Ohliger}\ and\ \citenamefont
  {Eisert}(2012)}]{Ohliger2012}%
  \BibitemOpen
  \bibfield  {author} {\bibinfo {author} {\bibfnamefont {M.}~\bibnamefont
  {Ohliger}}\ and\ \bibinfo {author} {\bibfnamefont {J.}~\bibnamefont
  {Eisert}},\ }\href {\doibase 10.1103/PhysRevA.85.062318} {\bibfield
  {journal} {\bibinfo  {journal} {Phys. Rev. A}\ }\textbf {\bibinfo {volume}
  {85}},\ \bibinfo {eid} {062318} (\bibinfo {year} {2012})}\BibitemShut
  {NoStop}%
\bibitem [{\citenamefont {Kot}\ \emph {et~al.}(2012)\citenamefont {Kot},
  \citenamefont {Grønbech-Jensen}, \citenamefont {Nielsen}, \citenamefont
  {Neergaard-Nielsen}, \citenamefont {Polzik},\ and\ \citenamefont
  {Sørensen}}]{Kot2012}%
  \BibitemOpen
  \bibfield  {author} {\bibinfo {author} {\bibfnamefont {E.}~\bibnamefont
  {Kot}}, \bibinfo {author} {\bibfnamefont {N.}~\bibnamefont
  {Grønbech-Jensen}}, \bibinfo {author} {\bibfnamefont {B.~M.}\ \bibnamefont
  {Nielsen}}, \bibinfo {author} {\bibfnamefont {J.~S.}\ \bibnamefont
  {Neergaard-Nielsen}}, \bibinfo {author} {\bibfnamefont {E.~S.}\ \bibnamefont
  {Polzik}}, \ and\ \bibinfo {author} {\bibfnamefont {A.~S.}\ \bibnamefont
  {Sørensen}},\ }\href {\doibase 10.1103/PhysRevLett.108.233601} {\bibfield
  {journal} {\bibinfo  {journal} {Phys. Rev. Lett.}\ }\textbf {\bibinfo
  {volume} {108}},\ \bibinfo {eid} {233601} (\bibinfo {year}
  {2012})}\BibitemShut {NoStop}%
\bibitem [{\citenamefont {Dowling}\ \emph {et~al.}(1994)\citenamefont
  {Dowling}, \citenamefont {Agarwal},\ and\ \citenamefont
  {Schleich}}]{Dowling1994}%
  \BibitemOpen
  \bibfield  {author} {\bibinfo {author} {\bibfnamefont {J.~P.}\ \bibnamefont
  {Dowling}}, \bibinfo {author} {\bibfnamefont {G.~S.}\ \bibnamefont
  {Agarwal}}, \ and\ \bibinfo {author} {\bibfnamefont {W.~P.}\ \bibnamefont
  {Schleich}},\ }\href {\doibase 10.1103/PhysRevA.49.4101} {\bibfield
  {journal} {\bibinfo  {journal} {Phys. Rev. A}\ }\textbf {\bibinfo {volume}
  {49}},\ \bibinfo {pages} {4101} (\bibinfo {year} {1994})}\BibitemShut
  {NoStop}%
\bibitem [{\citenamefont {Appel}\ \emph {et~al.}(2007)\citenamefont {Appel},
  \citenamefont {Hoffman}, \citenamefont {Figueroa},\ and\ \citenamefont
  {Lvovsky}}]{Appel2007}%
  \BibitemOpen
  \bibfield  {author} {\bibinfo {author} {\bibfnamefont {J.}~\bibnamefont
  {Appel}}, \bibinfo {author} {\bibfnamefont {D.}~\bibnamefont {Hoffman}},
  \bibinfo {author} {\bibfnamefont {E.}~\bibnamefont {Figueroa}}, \ and\
  \bibinfo {author} {\bibfnamefont {A.~I.}\ \bibnamefont {Lvovsky}},\ }\href
  {\doibase 10.1103/PhysRevA.75.035802} {\bibfield  {journal} {\bibinfo
  {journal} {Phys. Rev. A}\ }\textbf {\bibinfo {volume} {75}},\ \bibinfo
  {pages} {035802} (\bibinfo {year} {2007})}\BibitemShut {NoStop}%
\bibitem [{\citenamefont {Kiesel}\ \emph {et~al.}(2012)\citenamefont {Kiesel},
  \citenamefont {Vogel}, \citenamefont {Christensen}, \citenamefont {Béguin},
  \citenamefont {Appel},\ and\ \citenamefont {Polzik}}]{Kiesel2012}%
  \BibitemOpen
  \bibfield  {author} {\bibinfo {author} {\bibfnamefont {T.}~\bibnamefont
  {Kiesel}}, \bibinfo {author} {\bibfnamefont {W.}~\bibnamefont {Vogel}},
  \bibinfo {author} {\bibfnamefont {S.~L.}\ \bibnamefont {Christensen}},
  \bibinfo {author} {\bibfnamefont {J.-B.}\ \bibnamefont {Béguin}}, \bibinfo
  {author} {\bibfnamefont {J.}~\bibnamefont {Appel}}, \ and\ \bibinfo {author}
  {\bibfnamefont {E.~S.}\ \bibnamefont {Polzik}},\ }\href {\doibase
  10.1103/PhysRevA.86.042108} {\bibfield  {journal} {\bibinfo  {journal} {Phys.
  Rev. A}\ }\textbf {\bibinfo {volume} {86}},\ \bibinfo {pages} {042108}
  (\bibinfo {year} {2012})}\BibitemShut {NoStop}%
\bibitem [{\citenamefont {Laurat}\ \emph {et~al.}(2006)\citenamefont {Laurat},
  \citenamefont {de~Riedmatten}, \citenamefont {Felinto}, \citenamefont {Chou},
  \citenamefont {Schomburg},\ and\ \citenamefont {Kimble}}]{Laurat2006}%
  \BibitemOpen
  \bibfield  {author} {\bibinfo {author} {\bibfnamefont {J.}~\bibnamefont
  {Laurat}}, \bibinfo {author} {\bibfnamefont {H.}~\bibnamefont
  {de~Riedmatten}}, \bibinfo {author} {\bibfnamefont {D.}~\bibnamefont
  {Felinto}}, \bibinfo {author} {\bibfnamefont {C.-W.}\ \bibnamefont {Chou}},
  \bibinfo {author} {\bibfnamefont {E.~W.}\ \bibnamefont {Schomburg}}, \ and\
  \bibinfo {author} {\bibfnamefont {H.~J.}\ \bibnamefont {Kimble}},\ }\href
  {\doibase 10.1364/OE.14.006912} {\bibfield  {journal} {\bibinfo  {journal}
  {Opt. Express}\ }\textbf {\bibinfo {volume} {14}},\ \bibinfo {pages} {6912}
  (\bibinfo {year} {2006})}\BibitemShut {NoStop}%
\bibitem [{\citenamefont {Kuzmich}\ \emph {et~al.}(2003)\citenamefont
  {Kuzmich}, \citenamefont {Bowen}, \citenamefont {Boozer}, \citenamefont
  {Boca}, \citenamefont {Chou}, \citenamefont {Duan},\ and\ \citenamefont
  {Kimble}}]{Kuzmich2003}%
  \BibitemOpen
  \bibfield  {author} {\bibinfo {author} {\bibfnamefont {A.}~\bibnamefont
  {Kuzmich}}, \bibinfo {author} {\bibfnamefont {W.~P.}\ \bibnamefont {Bowen}},
  \bibinfo {author} {\bibfnamefont {A.~D.}\ \bibnamefont {Boozer}}, \bibinfo
  {author} {\bibfnamefont {A.}~\bibnamefont {Boca}}, \bibinfo {author}
  {\bibfnamefont {C.~W.}\ \bibnamefont {Chou}}, \bibinfo {author}
  {\bibfnamefont {L.-M.}\ \bibnamefont {Duan}}, \ and\ \bibinfo {author}
  {\bibfnamefont {H.~J.}\ \bibnamefont {Kimble}},\ }\href {\doibase
  10.1038/nature01714} {\bibfield  {journal} {\bibinfo  {journal} {Nature}\
  }\textbf {\bibinfo {volume} {423}},\ \bibinfo {pages} {731} (\bibinfo {year}
  {2003})}\BibitemShut {NoStop}%
\bibitem [{\citenamefont {Dubost}\ \emph {et~al.}(2012)\citenamefont {Dubost},
  \citenamefont {Koschorreck}, \citenamefont {Napolitano}, \citenamefont
  {Behbood}, \citenamefont {Sewell},\ and\ \citenamefont
  {Mitchell}}]{Dubost2012}%
  \BibitemOpen
  \bibfield  {author} {\bibinfo {author} {\bibfnamefont {B.}~\bibnamefont
  {Dubost}}, \bibinfo {author} {\bibfnamefont {M.}~\bibnamefont {Koschorreck}},
  \bibinfo {author} {\bibfnamefont {M.}~\bibnamefont {Napolitano}}, \bibinfo
  {author} {\bibfnamefont {N.}~\bibnamefont {Behbood}}, \bibinfo {author}
  {\bibfnamefont {R.~J.}\ \bibnamefont {Sewell}}, \ and\ \bibinfo {author}
  {\bibfnamefont {M.~W.}\ \bibnamefont {Mitchell}},\ }\href {\doibase
  10.1103/PhysRevLett.108.183602} {\bibfield  {journal} {\bibinfo  {journal}
  {Phys. Rev. Lett.}\ }\textbf {\bibinfo {volume} {108}},\ \bibinfo {eid}
  {183602} (\bibinfo {year} {2012})}\BibitemShut {NoStop}%
\bibitem [{\citenamefont {Schmied}\ and\ \citenamefont
  {Treutlein}(2011)}]{Schmied2011}%
  \BibitemOpen
  \bibfield  {author} {\bibinfo {author} {\bibfnamefont {R.}~\bibnamefont
  {Schmied}}\ and\ \bibinfo {author} {\bibfnamefont {P.}~\bibnamefont
  {Treutlein}},\ }\href {\doibase 10.1088/1367-2630/13/6/065019} {\bibfield
  {journal} {\bibinfo  {journal} {New. J. Phys.}\ }\textbf {\bibinfo {volume}
  {13}},\ \bibinfo {pages} {065019} (\bibinfo {year} {2011})}\BibitemShut
  {NoStop}%
\bibitem [{\citenamefont {Vetsch}\ \emph {et~al.}(2010)\citenamefont {Vetsch},
  \citenamefont {Reitz}, \citenamefont {Sagu{\'e}}, \citenamefont {Schmidt},
  \citenamefont {Dawkins},\ and\ \citenamefont {Rauschenbeutel}}]{Vetsch2010}%
  \BibitemOpen
  \bibfield  {author} {\bibinfo {author} {\bibfnamefont {E.}~\bibnamefont
  {Vetsch}}, \bibinfo {author} {\bibfnamefont {D.}~\bibnamefont {Reitz}},
  \bibinfo {author} {\bibfnamefont {G.}~\bibnamefont {Sagu{\'e}}}, \bibinfo
  {author} {\bibfnamefont {R.}~\bibnamefont {Schmidt}}, \bibinfo {author}
  {\bibfnamefont {S.~T.}\ \bibnamefont {Dawkins}}, \ and\ \bibinfo {author}
  {\bibfnamefont {A.}~\bibnamefont {Rauschenbeutel}},\ }\href {\doibase
  10.1103/PhysRevLett.104.203603} {\bibfield  {journal} {\bibinfo  {journal}
  {Phys. Rev. Lett.}\ }\textbf {\bibinfo {volume} {104}},\ \bibinfo {pages}
  {203603} (\bibinfo {year} {2010})}\BibitemShut {NoStop}%
\bibitem [{\citenamefont {Lacroûte}\ \emph {et~al.}(2012)\citenamefont
  {Lacroûte}, \citenamefont {Choi}, \citenamefont {Goban}, \citenamefont
  {Alton}, \citenamefont {Ding}, \citenamefont {Stern},\ and\ \citenamefont
  {Kimble}}]{Lacroute2012}%
  \BibitemOpen
  \bibfield  {author} {\bibinfo {author} {\bibfnamefont {C.}~\bibnamefont
  {Lacroûte}}, \bibinfo {author} {\bibfnamefont {K.~S.}\ \bibnamefont {Choi}},
  \bibinfo {author} {\bibfnamefont {A.}~\bibnamefont {Goban}}, \bibinfo
  {author} {\bibfnamefont {D.~J.}\ \bibnamefont {Alton}}, \bibinfo {author}
  {\bibfnamefont {D.}~\bibnamefont {Ding}}, \bibinfo {author} {\bibfnamefont
  {N.~P.}\ \bibnamefont {Stern}}, \ and\ \bibinfo {author} {\bibfnamefont
  {H.~J.}\ \bibnamefont {Kimble}},\ }\href {\doibase
  10.1088/1367-2630/14/2/023056} {\bibfield  {journal} {\bibinfo  {journal}
  {New. J. Phys.}\ }\textbf {\bibinfo {volume} {14}},\ \bibinfo {pages}
  {023056} (\bibinfo {year} {2012})}\BibitemShut {NoStop}%
\bibitem [{\citenamefont {Zhang}\ \emph {et~al.}(2012)\citenamefont {Zhang},
  \citenamefont {McConnell}, \citenamefont {Ćuk}, \citenamefont {Lin},
  \citenamefont {Schleier-Smith}, \citenamefont {Leroux},\ and\ \citenamefont
  {Vuletić}}]{Zhang2012}%
  \BibitemOpen
  \bibfield  {author} {\bibinfo {author} {\bibfnamefont {H.}~\bibnamefont
  {Zhang}}, \bibinfo {author} {\bibfnamefont {R.}~\bibnamefont {McConnell}},
  \bibinfo {author} {\bibfnamefont {S.}~\bibnamefont {Ćuk}}, \bibinfo {author}
  {\bibfnamefont {Q.}~\bibnamefont {Lin}}, \bibinfo {author} {\bibfnamefont
  {M.~H.}\ \bibnamefont {Schleier-Smith}}, \bibinfo {author} {\bibfnamefont
  {I.~D.}\ \bibnamefont {Leroux}}, \ and\ \bibinfo {author} {\bibfnamefont
  {V.}~\bibnamefont {Vuletić}},\ }\href {\doibase
  10.1103/PhysRevLett.109.133603} {\bibfield  {journal} {\bibinfo  {journal}
  {Phys. Rev. Lett.}\ }\textbf {\bibinfo {volume} {109}},\ \bibinfo {eid}
  {133603} (\bibinfo {year} {2012})}\BibitemShut {NoStop}%
\bibitem [{\citenamefont {{Colombe}}\ \emph {et~al.}(2007)\citenamefont
  {{Colombe}}, \citenamefont {{Steinmetz}}, \citenamefont {{Dubois}},
  \citenamefont {{Linke}}, \citenamefont {{Hunger}},\ and\ \citenamefont
  {{Reichel}}}]{Colombe2007}%
  \BibitemOpen
  \bibfield  {author} {\bibinfo {author} {\bibfnamefont {Y.}~\bibnamefont
  {{Colombe}}}, \bibinfo {author} {\bibfnamefont {T.}~\bibnamefont
  {{Steinmetz}}}, \bibinfo {author} {\bibfnamefont {G.}~\bibnamefont
  {{Dubois}}}, \bibinfo {author} {\bibfnamefont {F.}~\bibnamefont {{Linke}}},
  \bibinfo {author} {\bibfnamefont {D.}~\bibnamefont {{Hunger}}}, \ and\
  \bibinfo {author} {\bibfnamefont {J.}~\bibnamefont {{Reichel}}},\ }\href
  {\doibase 10.1038/nature06331} {\bibfield  {journal} {\bibinfo  {journal}
  {Nature}\ }\textbf {\bibinfo {volume} {450}},\ \bibinfo {pages} {272}
  (\bibinfo {year} {2007})}\BibitemShut {NoStop}%
\bibitem [{\citenamefont {Thompson}\ \emph {et~al.}(2013)\citenamefont
  {Thompson}, \citenamefont {Tiecke}, \citenamefont {de~Leon}, \citenamefont
  {Feist}, \citenamefont {Akimov}, \citenamefont {Gullans}, \citenamefont
  {Zibrov}, \citenamefont {Vuleti{\'c}},\ and\ \citenamefont
  {Lukin}}]{Thompson2013}%
  \BibitemOpen
  \bibfield  {author} {\bibinfo {author} {\bibfnamefont {J.~D.}\ \bibnamefont
  {Thompson}}, \bibinfo {author} {\bibfnamefont {T.~G.}\ \bibnamefont
  {Tiecke}}, \bibinfo {author} {\bibfnamefont {N.~P.}\ \bibnamefont {de~Leon}},
  \bibinfo {author} {\bibfnamefont {J.}~\bibnamefont {Feist}}, \bibinfo
  {author} {\bibfnamefont {A.~V.}\ \bibnamefont {Akimov}}, \bibinfo {author}
  {\bibfnamefont {M.}~\bibnamefont {Gullans}}, \bibinfo {author} {\bibfnamefont
  {A.~S.}\ \bibnamefont {Zibrov}}, \bibinfo {author} {\bibfnamefont
  {V.}~\bibnamefont {Vuleti{\'c}}}, \ and\ \bibinfo {author} {\bibfnamefont
  {M.~D.}\ \bibnamefont {Lukin}},\ }\href {\doibase 10.1126/science.1237125}
  {\bibfield  {journal} {\bibinfo  {journal} {Science}\ }\textbf {\bibinfo
  {volume} {340}},\ \bibinfo {pages} {1202} (\bibinfo {year}
  {2013})}\BibitemShut {NoStop}%
\bibitem [{\citenamefont {{Goban}}\ \emph {et~al.}(2013)\citenamefont
  {{Goban}}, \citenamefont {{Hung}}, \citenamefont {{Yu}}, \citenamefont
  {{Hood}}, \citenamefont {{Muniz}}, \citenamefont {{Lee}}, \citenamefont
  {{Martin}}, \citenamefont {{McClung}}, \citenamefont {{Choi}}, \citenamefont
  {{Chang}}, \citenamefont {{Painter}},\ and\ \citenamefont
  {{Kimble}}}]{Goban2013}%
  \BibitemOpen
  \bibfield  {author} {\bibinfo {author} {\bibfnamefont {A.}~\bibnamefont
  {{Goban}}}, \bibinfo {author} {\bibfnamefont {C.-L.}\ \bibnamefont {{Hung}}},
  \bibinfo {author} {\bibfnamefont {S.-P.}\ \bibnamefont {{Yu}}}, \bibinfo
  {author} {\bibfnamefont {J.~D.}\ \bibnamefont {{Hood}}}, \bibinfo {author}
  {\bibfnamefont {J.~A.}\ \bibnamefont {{Muniz}}}, \bibinfo {author}
  {\bibfnamefont {J.~H.}\ \bibnamefont {{Lee}}}, \bibinfo {author}
  {\bibfnamefont {M.~J.}\ \bibnamefont {{Martin}}}, \bibinfo {author}
  {\bibfnamefont {A.~C.}\ \bibnamefont {{McClung}}}, \bibinfo {author}
  {\bibfnamefont {K.~S.}\ \bibnamefont {{Choi}}}, \bibinfo {author}
  {\bibfnamefont {D.~E.}\ \bibnamefont {{Chang}}}, \bibinfo {author}
  {\bibfnamefont {O.}~\bibnamefont {{Painter}}}, \ and\ \bibinfo {author}
  {\bibfnamefont {H.~J.}\ \bibnamefont {{Kimble}}},\ }\href@noop {} {\bibfield
  {journal} {\bibinfo  {journal} {ArXiv e-prints}\ } (\bibinfo {year}
  {2013})},\ \Eprint {http://arxiv.org/abs/1312.3446} {arXiv:1312.3446}
  \BibitemShut {NoStop}%
\bibitem [{\citenamefont {Simon}\ and\ \citenamefont
  {Polzik}(2011)}]{Simon2011}%
  \BibitemOpen
  \bibfield  {author} {\bibinfo {author} {\bibfnamefont {C.}~\bibnamefont
  {Simon}}\ and\ \bibinfo {author} {\bibfnamefont {E.~S.}\ \bibnamefont
  {Polzik}},\ }\href {\doibase 10.1103/PhysRevA.83.040101} {\bibfield
  {journal} {\bibinfo  {journal} {Phys. Rev. A}\ }\textbf {\bibinfo {volume}
  {83}},\ \bibinfo {eid} {040101} (\bibinfo {year} {2011})}\BibitemShut
  {NoStop}%
\bibitem [{\citenamefont {Lee}(1990)}]{Lee1990}%
  \BibitemOpen
  \bibfield  {author} {\bibinfo {author} {\bibfnamefont {C.~T.}\ \bibnamefont
  {Lee}},\ }\href {\doibase 10.1103/PhysRevA.42.1608} {\bibfield  {journal}
  {\bibinfo  {journal} {Phys. Rev. A}\ }\textbf {\bibinfo {volume} {42}},\
  \bibinfo {pages} {1608} (\bibinfo {year} {1990})}\BibitemShut {NoStop}%
\end{thebibliography}%

\end{document}